\documentclass[structabstract]{aa}
\usepackage{pdfpages}
\usepackage{subfig}
\usepackage{graphicx}
\usepackage{txfonts}
\usepackage{natbib}
\usepackage[english]{babel}
\bibpunct{(}{)}{;}{a}{}{,}
\begin{document}
\title{A new interferometric study of four exoplanet host stars\,: $\theta$~Cygni, 14~Andromedae, $\upsilon$ Andromedae and 42~Draconis.
\thanks{Based on interferometric observations with the VEGA/CHARA instrument.}
}

\author{R. Ligi\inst{1}, D. Mourard\inst{1}, A.M. Lagrange\inst{2}, K. Perraut\inst{2}, T. Boyajian\inst{4}\thanks{Hubble Fellow}, Ph. B\'{e}rio\inst{1}, N. Nardetto\inst{1}, I. Tallon-Bosc\inst{3}, H. McAlister\inst{4,5}, T. ten~Brummelaar\inst{4}, S. Ridgway\inst{6}, J. Sturmann\inst{4}, L. Sturmann\inst{4}, N. Turner\inst{4}, C. Farrington\inst{4} and P.J. Goldfinger\inst{4}}
\institute{Laboratoire Lagrange, UMR 7293 UNS-CNRS-OCA, Boulevard de l'Observatoire, B.P. 4229 F, 06304 NICE Cedex 4, France.
   \and
   UJF-Grenoble1/CNRS-INSU, Institut de Plan\'{e}tologie et d'Astrophysique de Grenoble, UMR 5274, Grenoble, F-38041, France
   \and
   Universit\'{e}� de Lyon, 69003 Lyon\,: Universit\'{e}� Lyon 1, Observatoire de Lyon, 9 Avenue Charles Andr\'{e}�, 69230 Saint-Genis Laval\,:
CNRS, UMR 5574, Centre de Recherche Astrophysique de Lyon\,: Ecole Normale Sup\`{e}rieure de Lyon, 69007 Lyon, France
   \and
   Georgia State University, P.O. Box 3969, Atlanta GA 30302-3969, USA
   \and
   CHARA Array, Mount Wilson Observatory, 91023 Mount Wilson CA, USA
   \and
    National Optical Astronomy Observatory, PO Box 26732, Tucson, AZ 85726, USA}

\date{Received ...; accepted ...}
\abstract
{Since the discovery of the first exoplanet in 1995 around a solar-type star, the interest in exoplanetary systems has kept increasing. Studying exoplanet host stars is of the utmost importance to establish the link between the presence of exoplanets around various types of stars and to understand the respective evolution of stars and exoplanets.}
{Using the limb-darkened diameter (LDD) obtained from interferometric data, we determine the fundamental parameters of four exoplanet host stars. We are particularly interested in the F4 main-sequence star, $\theta$~Cyg, for which Kepler has recently revealed solar-like oscillations that are unexpected for this type of star. Furthermore, recent photometric and spectroscopic measurements with SOPHIE and ELODIE (OHP) show evidence of a quasi-periodic radial velocity of $\sim$ 150 days. Models of this periodic change in radial velocity predict either a complex planetary system orbiting the star, or a new and unidentified stellar pulsation mode.}
{We performed interferometric observations of $\theta$~Cyg, 14~Andromedae, $\upsilon$ Andromedae and 42~Draconis for two years with VEGA/CHARA (Mount Wilson, California) in several three-telescope configurations. We measured accurate limb darkened diameters and derived their radius, mass and temperature using empirical laws.}
{We obtain new accurate fundamental parameters for stars 14~And, $\upsilon$ And and 42~Dra. We also obtained limb darkened diameters with a minimum precision of $\sim1,3\%$, leading to minimum planet masses of $M \sin i = 5.33\pm0.57$, $0.62\pm0.09$ and $3.79\pm0.29$ $M_{\rm Jup}$ for 14~And~b, $\upsilon$~And~b and 42~Dra~b, respectively.
The interferometric measurements of $\theta$~Cyg show a significant diameter variability that remains unexplained up to now. We propose that the presence of these discrepancies in the interferometric data is caused by either an intrinsic variation of the star or an unknown close companion orbiting around it.}

{}

   \authorrunning{R. Ligi et al.}
   \titlerunning{A new interferometric study of four exoplanet host stars\,: $\theta$~Cyg, 14~And, $\upsilon$ And and 42~Dra.}
   \maketitle{}

\section{Introduction}
\label{sect:Intro}

Many techniques have been developed during the past decade to enable the discovery of exoplanets. The radial velocity method, based on the reflex motion of the host star, is one of the most successful of these and has to date enabled the discovery of 535 planetary systems\footnote{As of December 23, 2011 \citep{encyclopedia}}.
Most of these planets were found orbiting slowly rotating stars, late-type stars, or A giants. While A and F main sequence stars were usually avoided because of their high $v\sin i$, a survey of A and F main sequence stars was nonetheless recently undertaken using a specialized analysis method to look for planets around these stars, and planets were indeed found around a few F stars \citep{Lagrange2009}. Unfortunately, the possible planet configurations fitting the radial velocity (RV) data were found to be dynamically unstable. To resolve this problem it is important to better understand the link between the presence and mass of exoplanets, the host star parameters, and the separation of the planet and host star.

Interferometric data are now able to bring additional information to bear on stellar variability and its contribution to noise in the radial velocity measurements, and can help to directly determine many of the fundamental parameters of the host stars with an accuracy of about $5 \%$ see for example \cite[see for example][]{baines,55cnc}. This is not only very important for deriving accurate radii for transiting planets, but also for RV planets. Understanding the link between the presence and nature of exoplanets and the fundamental parameters of the star requires sampling a large number of targets. We have started a survey with VEGA (Visible spEctroGraph and polArimeter) \citep{Vega1}, a visible spectro-interferometer located on the CHARA (Center for High Angular Resolution Astronomy) array at Mount Wilson, California \citep{chara}, to measure all currently accessible exoplanets stars, i.e. almost 40 targets. To build this sample, we first selected the exoplanet host stars listed in Schneider's catalog \citep{encyclopedia}. Those stars have to be observable by VEGA, therefore we sorted out those that had a magnitude smaller than $6.5$ in the V- and in the K band, and a declination higher than $-30^{\circ}$. 
Knowing the error on the squared visibility allowed by VEGA at medium resolution ($\simeq 2\%$), we can estimate the maximum and the minimum diameters for which we can obtain an accuracy of $\simeq 2\%$ taking into account the maximum and minimum baselines. We consider this accuracy as the minimum allowed to obtain sufficiently good informations on fundamental parameters of the stars and planets. Diameters included between $0.3$ and $3$ milliseconds of arc (mas) are sufficiently resolved to achieve this accuracy. We finally found 40 stars whose planets were discovered with the transit or RV techniques.

Interferometry is complementary to the transit method or RV measurements in determining exoplanet parameters. For instance, the transit method allows determining the exoplanet radius, while the RV method is used to detect the minimum mass. The main goal of these observations is to directly constrain these parameters, and to study the impact of stellar noise sources (e.g., spots, limb darkening) applied to these observing methods. In the long term, the results will be compared to a catalog   of limb darkening laws from 3D hydro-­dynamical modeling and radiative transfer. Thus, we will be able to create a catalog of measured angular diameters, and derive revised surface brightness relationships.  

From October to December 2011, we obtained data on three stars of our sample\,: 14~And, $\upsilon$ And and 42~Dra, while a fourth star $\theta$~Cyg was observed over a longer period, from June 2010 to November 2011. We found that while the first three stars yield stable and repeatable results, there are  discrepancies in the results of $\theta$~Cyg, forcing us to study this system more carefully. New and unexplained RV variations recorded with SOPHIE and ELODIE at the Observatoire de Haute-Provence \citep{13cygDesort} provided a first clue that this star hosts either a complex planetary system, undergoes hitherto unknown variations, or has a hidden companion.

After a short introduction to the basics of interferometry, we describe in Section~\ref{sect:exoplanets} the observations made of 14~And, $\upsilon$ And and 42~Dra during the year 2011 and derive the star and planet fundamental parameters.  We then compare these values to those found in the literature. In Section~\ref{sect:observations}, we present the observations of $\theta$~Cyg made during the last two years. We discuss the fundamental parameters we derived for this target in Section~\ref{sect:parameters}, and compare them with the previously known parameters for this star (see Table~\ref{tab:table1}). We then discuss the variation of the angular diameter of $\theta$~Cyg in Section~\ref{sect:discussion} and some possible explanations of this variability.

\section{Observations with VEGA/CHARA}
\label{sect:interferometry}

\subsection{VEGA/CHARA and visibility determination}
\label{subsect:visibility}

\begin{figure}
	\includegraphics[width=8.5 cm]{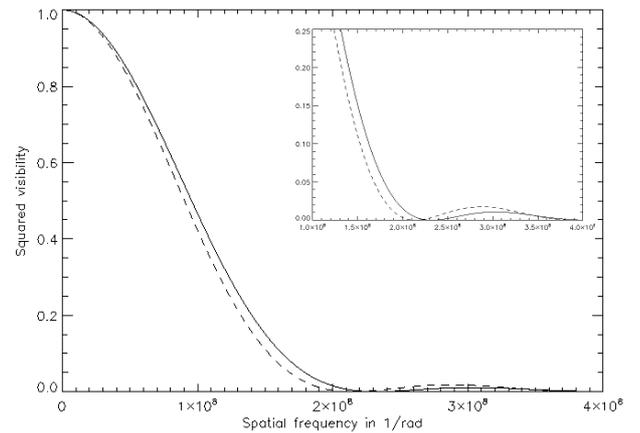}
	\caption{Squared visibility of a uniform disk (solid line) and of a limb-darkened disk (dashed line) for a star of $1.17$ mas of diameter, a wavelength of 720 nm and a baseline ranging from ~$0$ to $330$ m. The LDD is sensitive close to the zero and in the second lobe of visibility, where it is higher than for a UD.}
	\label{fig:ComparaisonLDUD}
\end{figure}

The CHARA array hosts six one-meter telescopes arranged in a Y shape that are oriented to the east (E1 and E2), south (S1 and S2) and west (W1 and W2). The baselines range between $34$ and $331$ m and permit a wide range of orientations. 
VEGA is a spectro-interferometer working in the visible wavelengths at different spectral resolutions\,: $6000$ and $30000$. Thus, it permits the recombination of two, three or four telescopes, and a maximum angular resolution of $\simeq~0.3$ mas.
Interferometry is a high angular resolution technique allowing one to study the spatial brightness distribution of celestial objects through measuring their spatial frequencies. By measuring the fringe contrast, also called visibility, one is able to determine the size of stars, thanks to the van Cittert-Zernike theorem \citep{born&wolf}. The simplest representation of a star is a uniform disk (UD) of angular diameter $\theta_{\rm UD}$. The corresponding visibility function is given by

\begin{equation}
V^{2} = \left| \frac{2J_{1}(x)}{x}\right| ^{2},
\end{equation} 

where $J_{1}(x)$ is the first-order Bessel function and $x = \pi B \theta_{UD} \lambda^{-1}$. $B$ represents the length of the projected baseline, $\lambda$ the wavelength of the observation. However, stars are not uniformly bright\,: a better representation of the surface brightness is the limb-darkened disk (LDD). The main differences between the two profiles arise close to the zero of visibility and in the second lobe, as shown in Figure~\ref{fig:ComparaisonLDUD}.

The LDD is conventionally described by the function $I_{\rm \lambda}[\mu]$, where $\mu$ is the cosine between the normal to the surface at that point and the line of sight from the star to the observer and $u_{\rm \lambda}$ the limb darkening coefficient \citep{Hanbury}\,:

\begin{equation}
I_{\lambda}[\mu] = I_{\lambda}[1][1-u_{\lambda}(1-\mu)].
\end{equation}

A good approximation of $\theta_{\rm LD}$ is given by

\begin{equation}
\theta_{LD}[\lambda] = \theta_{UD}[\lambda] \times \left[\frac{1-u_{\lambda}/3}{1-7u_{\lambda}/15}\right]^{1/2},
\end{equation}

\citep{Hanbury}.
 
The \cite{claret2011} coefficients are listed in tables and depend on the effective temperature and the $\log(g)$. We calculated that in our observing conditions, a difference of $10 \%$ on the coefficients leads to a difference of $0.65 \%$ on the LDD and of $0.33 \%$ on the $T_{\rm eff}$. Using approximated coefficients is then of negligible consequence on the final parameters' values.

Because we performed 3T observations, we obtained three calibrated squared visibility points for each observation in the observed spectral band. The systematic and  statistical errors were calculated for each data point. The systematic error accounts for the influence of the estimated error on the angular diameter of the calibrators. In almost all cases, the systematic error is negligible compared to the statistical one by a factor 10, because our diameters are small \citep{Vega1}. The statistical error takes into account the instrumental variations, the variations of atmospheric conditions (seeing), and vibrations of the telescopes or the delay lines. It is measured when we estimate the noise and the error on the noise.

\subsection{Determination of the fundamental parameters}

We used empirical relations to derive the fundamental parameters of the stellar and planet components.
From the LDD ($\theta_{\rm LD}$) expressed in mas and the parallax ($\pi$) given in second of arc, we calculated the star's linear radius ($R$) and mass in the following manner. Using a simple Monte Carlo simulation, we obtain a correct estimate of the radius and its error\,:

\begin{equation}
\label{eq:rayon}
R\pm\delta R \emph{$(R_{\odot})$}= \frac{\theta_{LD}\pm\delta\theta_{LD}}{9.305\times(\pi\pm\delta\pi)}.
\label{eq:radius}
\end{equation}

We then use the gravitational acceleration relation to estimate the mass\,:

\begin{equation}
\Vert\overrightarrow{g}\Vert = GM / R^{2},
\label{eq:rm1}
\end{equation}

where $G$ is the gravitational constant. The modulus of $g$ is given in Table~\ref{tab:table5}. The error of the mass estimate is dominated by the uncertainty in parallax.
We also estimated the effective temperature using the black body law and the luminosities ($L$) shown in Table~\ref{tab:table5}\,:

\begin{equation}
L = 4 \pi R^{2} \sigma T_{\it{eff}}^{4}.
\label{eq:Teff}
\end{equation}

Starting from the stellar masses, we use the mass function to determine the exoplanet masses and estimate its error by performing a Monte Carlo test\,:

\begin{equation}
	\label{eq:massfunction}
f(m) = \frac{M_{pl}^{3}\sin(i)^{3}}{(M_{\ast}+M_{pl})^{2}},
\end{equation}

where $M_{\rm pl}$ and $M_{\ast}$ are the planet and stellar masses respectively. The results of the calculated planet masses are given in Table~\ref{tab:table10}.

Given that $M_{\rm pl} \ll M_{\ast}$ and using Kepler's third law, we can write\,

\begin{equation}
 \label{eq:Mjup}
M_{pl} \sin(i) = \frac{M_{\ast}^{2/3}P^{1/3}K(1-e^{2})^{1/2}}{(2\pi G)^{1/3}},
\end{equation}

where $\textit{K}$ is the velocity semi-amplitude and $\textit{e}$ the planet eccentricity.

\section{3T measurements of 14~And, $\upsilon$ And and 42~Dra}
\label{sect:exoplanets}

\subsection{VEGA observations}
In 2011, we observed two giant stars, 42~Dra (K1.5III\,: \cite{42Dra}) and 14~And (K0III\,: \cite{14And}), and one main-sequence star, $\upsilon$ And (F9V\,: \cite{uAnd}). The observations provided measurements close to the zero or up to the second lobe of squared visibility.

\begin{table}[h]
	\caption{Coordinates and parameters of the three host stars 14~And, $\upsilon$ And and 42~Dra. References from $^{(a)}$\cite{14And}\,: $^{(b)}$\cite{uAnd}\,: $^{(c)}$\cite{42Dra}\,: $^{(d)}$\cite{leeuwen}\,: $^{(e)}$\cite{Butler}\,: $^{(f)}$\cite{Baines2010}.}
	\label{tab:table5}
    \centering
	\begin{tabular}{l l l l l}
	
\hline\hline
Parameter & 14~And & $\upsilon$ And & 42~Dra \\
\hline
\centering
RA (J2000) & 23:31:17.4 & 01:36:47.8 & 18:25:59.14 \\
Dec (J2000) & $+39^{\circ}$14'10" & $+41^{\circ}$24'20" & $+65^{\circ}$33'49" \\
Stellar type & K0III$^{(a)}$ & F9V & K1.5III$^{(c)}$ \\
V mag  & 5.225 & 4.10 & 4.833$^{(c)}$ \\
K mag  & 2.331 & 2.86 & 2.085 \\
$M_{V}$ & 0.67$^{(a)}$ & 3.44$\pm$0.02$^{(b)}$ & -0.09$\pm$0.04$^{(c)}$\\
$v\,\sin i$ [km/s] & 2.60$^{(a)}$ & 9.5$\pm$0.8$^{(b)}$ & \\
$T_{\rm eff}$ [K] & 4813$\pm$20$^{(a)}$ & 6107$\pm$80$^{(b)}$ & 4200$\pm$70$^{(c)}$\\
Parallax [mas] & 12.63$\pm$0.27$^{(d)}$ & 74.12$\pm$0.19$^{(d)}$ & 10.36$\pm$0.20$^{(d)}$ \\
Mass [$M_\odot$] & 2.2$^{+0.1}_{-0.2}$ $^{(a)}$ & 1.27$\pm$0.06$^{(b)}$ & 0.98$\pm$0.05$^{(c)}$ \\
log g & 2.63$\pm$0.07$^{(a)}$ & 4.01$\pm$0.1$^{(b)}$ & 1.71$\pm$0.05$^{(c)}$ \\
$[Fe/H]$ & -0.24$\pm$0.03$^{(a)}$ & 0.09$\pm$0.006$^{(b)}$ & -0.46$\pm$0.05$^{(c)}$ \\
L [$L_\odot$] & 58$^{(a)}$ & 3$^{(e)}$ & 149.7$\pm$15.3$^{(f)}$\\
\hline
\end{tabular}
\end{table}

14~And (HD221345, HIP116076, HR8930) hosts one exoplanet of minimum mass $M_{2} \sin i = 4.8 M_{\rm J}$ discovered in 2008, and it has also been shown that this star does not exhibit measurable chromospheric activity \citep{14And}. The general properties of this star are given in Table~\ref{tab:table5}.

$\upsilon$ And (HD9826,	HIP7513, HR458) is a bright F star that has undergone numerous spectroscopic investigations \citep[][and references therein]{uAnd}. Four exoplanets are known to orbit around it\,: they were discovered between 1996 and 2010 \citep{encyclopedia, Butler, Lowrance2002, Curiel}.

42~Dra (HD170693, HIP90344, HR6945) is an intermediate-mass giant star around which a $3.88\pm0.85 M_{\rm J}$ exoplanet has recently been discovered \citep{42Dra}.

	\begin{table}[h]
	\caption{Parameters of the calibrators used for 14~And, $\upsilon$ And and 42~Dra. The value of the equivalent uniform disk $\theta_{\rm UD}$ is given at 700nm \citep{searchcal}.}
	\label{tab:table6}
	\begin{tabular}{llllll}
\hline\hline
$\sharp$ & Name & Spectral Type & $m_{V}$ & $m_{K}$ & $\theta_{\rm UD} [mas]$\\
\hline
	1 & HD 211211 & A2Vnn & 5.71 & 5.63 & 0.20$\pm$0.01 \\
	2 & HD 1439 & A0IV & 5.87 & 5.86 & 0.18$\pm$0.01 \\
	3 & HD 14212 & A1V & 5.31 & 5.27 & 0.24$\pm$0.02 \\
	4 & HD 187340 & A2III & 5.90 & 5.71 & 0.21$\pm$0.02 \\
\hline
	\end{tabular}
	\end{table}

Observations of these three exoplanet host stars were made in October and November 2011 with the E1E2W2 triplet. The data processing and the results analysis were presented in Subsection~\ref{subsect:visibility}. We used the calibrators HD211211 ($\textit{cal1}$) and HD1439 ($\textit{cal2}$) for 14~And, HD14212 ($\textit{cal3}$) for $\upsilon$ And and HD187340 ($\textit{cal4}$) for 42~Dra (Table~\ref{tab:table6}). They were found using the SearchCal utility\footnote{Available at http://www.jmmc.fr/searchcal} developed by the JMMC \citep{searchcal}. It gives, among other parameters, the stellar magnitude in the V and K bands, the spectral type, and also an estimate of the angular diameter along with the corresponding error. Angular diameters are determined by surface-brightness versus color-index relationships. We used the $V/(V-K)$ polynomial relation given by \cite{searchcal}. Its accuracy of $7 \%$ is the highest concerning the color-index polynomial fits.
We mainly observed with the three telescope (3T) configuration, but sometimes the conditions only allowed for 2T measurements (Table~\ref{tab:table7}). VEGA data are recorded as blocks of 1000 frames each of 15 milliseconds of exposure time. The observations of the targets were 30 minutes long (60 blocks), and those of the calibrators were 10 to 20 minutes long (20 or 40 blocks).
The data were recorded at medium spectral resolution ($R=6000$) and the data processing used 15 nm wide channels in the continuum of the red spectrum. We alternated the calibrators and target using the standard sequence $\textit{Cal-Target-Cal}$, which provides a better estimate of the transfer function during the observations of the target. We know \citep{Vega1} that, under correct seeing conditions, the transfer function of VEGA/CHARA is stable at the level of $2 \%$ for more than one hour. This has been checked in all our data set, and bad sequences were removed. We used the CLIMB beam combiner operating in the near-infrared as a 3T fringe tracker \citep{climb} to stabilize the optical path differences during the long integrations.

\begin{table}
	\caption{Journal of the observations of 14~And, $\upsilon$ And and 42~Dra. RJD is the reduced Julian day. The projected baseline is given by baseline (in meters) and PA in degree. $V^{2}$ is the calibrated squared visibility, the error of the squared visibility includes the statistical and systematic errors. All measurements use a band of 15nm around 707.5nm, except for the last observation of 42~Dra, which was centered around 732.5nm. In most cases ($^{\ast}$), CLIMB data in K band are also available.}
	\label{tab:table7}
	\begin{tabular}{llllllll}
\hline\hline
Star & RJD & Seq & Base & PA & $V^{2}$\\
\hline
14~And & 55855.4 & 1T1 & 66 & -123.4 & 0.306$\pm$0.022 \\ 
 	& & & 222 & -118.7 & 0.004$\pm$0.032 \\
 	& 55849.68$^{\ast}$ &  1T1 & 104 & 109.1 & 0.047$\pm$0.015 \\ 
 	& & & 153 & -108.2 & 0.012$\pm$0.016 \\
 	& & & 244 & -93.2 & 0.008$\pm$0.017  \\
 	& 55847.77$^{\ast}$ & 1T1 & 65 & -134.1 & 0.321$\pm$0.016 \\ 
 	& & & 154 & -127.2 & 0.020$\pm$0.008 \\
 	& 55847.72$^{\ast}$ & 1T2 & 66 & -122.9 & 0.420$\pm$0.022 \\ 
 	& & & 156 & -118.1 & 0.039$\pm$0.012 \\

$\upsilon$ And & 55883.74$^{\ast}$ & 3T3 & 66 & -131.3 & 0.384$\pm$0.020 \\ 
 	& & & 156 & -124.46 & 0.007$\pm$0.009  \\
 	& & & 221 &  -126.4 & 0.007$\pm$0.009  \\
 	& 55855.69$^{\ast}$ & 3T3 & 92 & 131 & 0.277$\pm$0.011 \\ 
 	& 55855.72$^{\ast}$ & 3T3 & 95.9 & 124.2 & 0.245$\pm$0.009 \\ 
 	& 55855.85$^{\ast}$ & 3T & 107 & 89.9 & 0.226$\pm$0.012 \\ 
	& 55854.78$^{\ast}$ & 3T3 & 156 & -120.2 & 0.437$\pm$0.027 \\ 
	& & & 151 & -113.5 & 0.000$\pm$0.007 \\
	& & & 221 & -115.5 & 0.023$\pm$0.011 \\

42~Dra & 55883.63$^{\ast}$& 4T4 & 66 & 169.5 & 0.100$\pm$0.015 \\ 
 	& 55854.63$^{\ast}$ & 4T4 & 66 & -164.9 & 0.086$\pm$0.007 \\ 
 	& 55854.63$^{\ast}$ & 4T4 & 66 & -164.9 & 0.111$\pm$0.009 \\ 
 	& & & 156 & -159.0 & 0.000$\pm$0.006 \\
 	& & & 222 & -160.7 & 0.006$\pm$0.011 \\

\hline
	\end{tabular}
\end{table}

\subsection{Fundamental parameters of stars and planets}

Because our data sets are covering many frequencies in the second lobe of the visibility function, we decided to fix the LDD coefficient and to adjust the diameter only. We used \cite{claret2011} tables. 

\begin{figure}
	\includegraphics[width=8.5 cm]{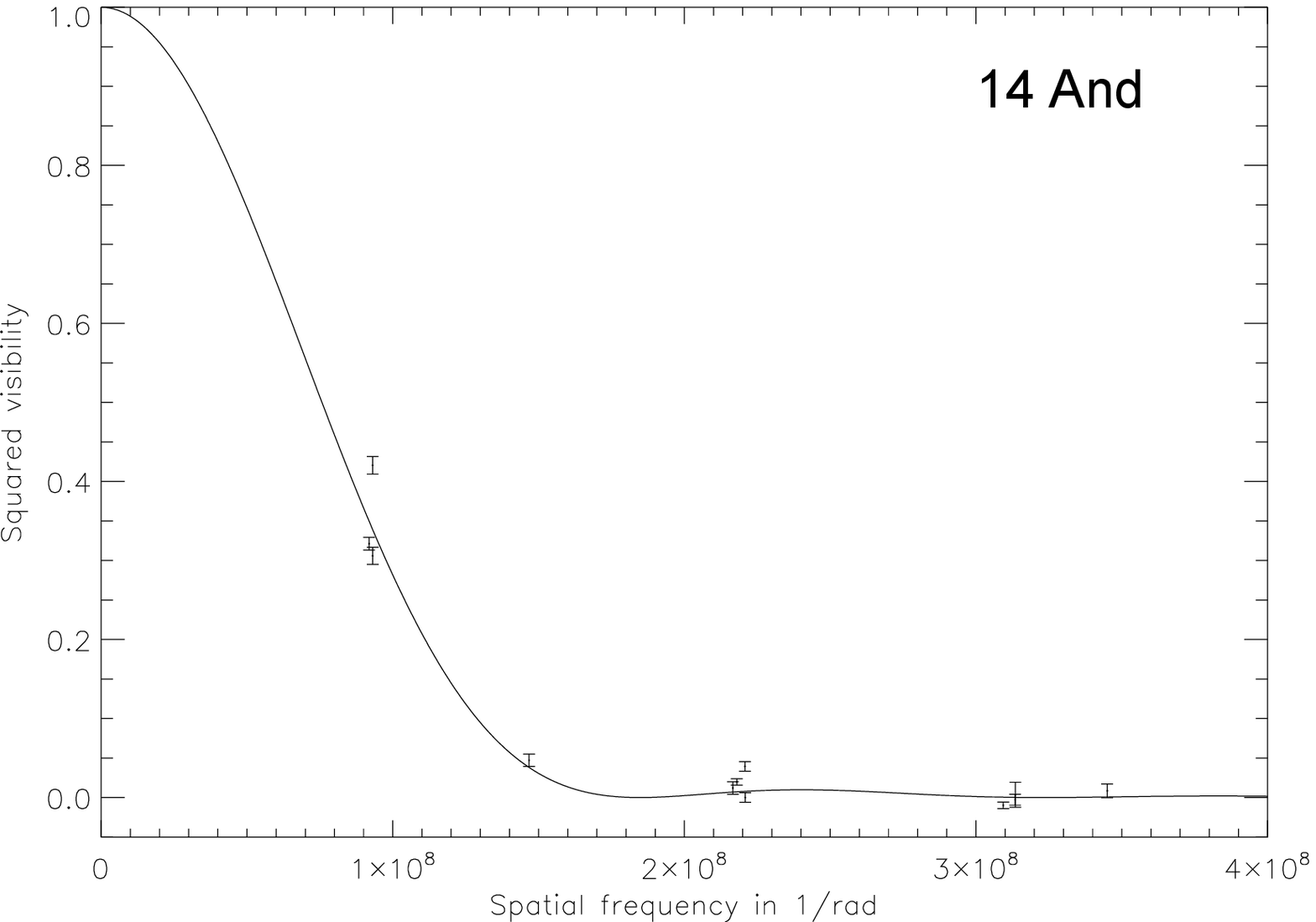}
	\includegraphics[width=8.5 cm]{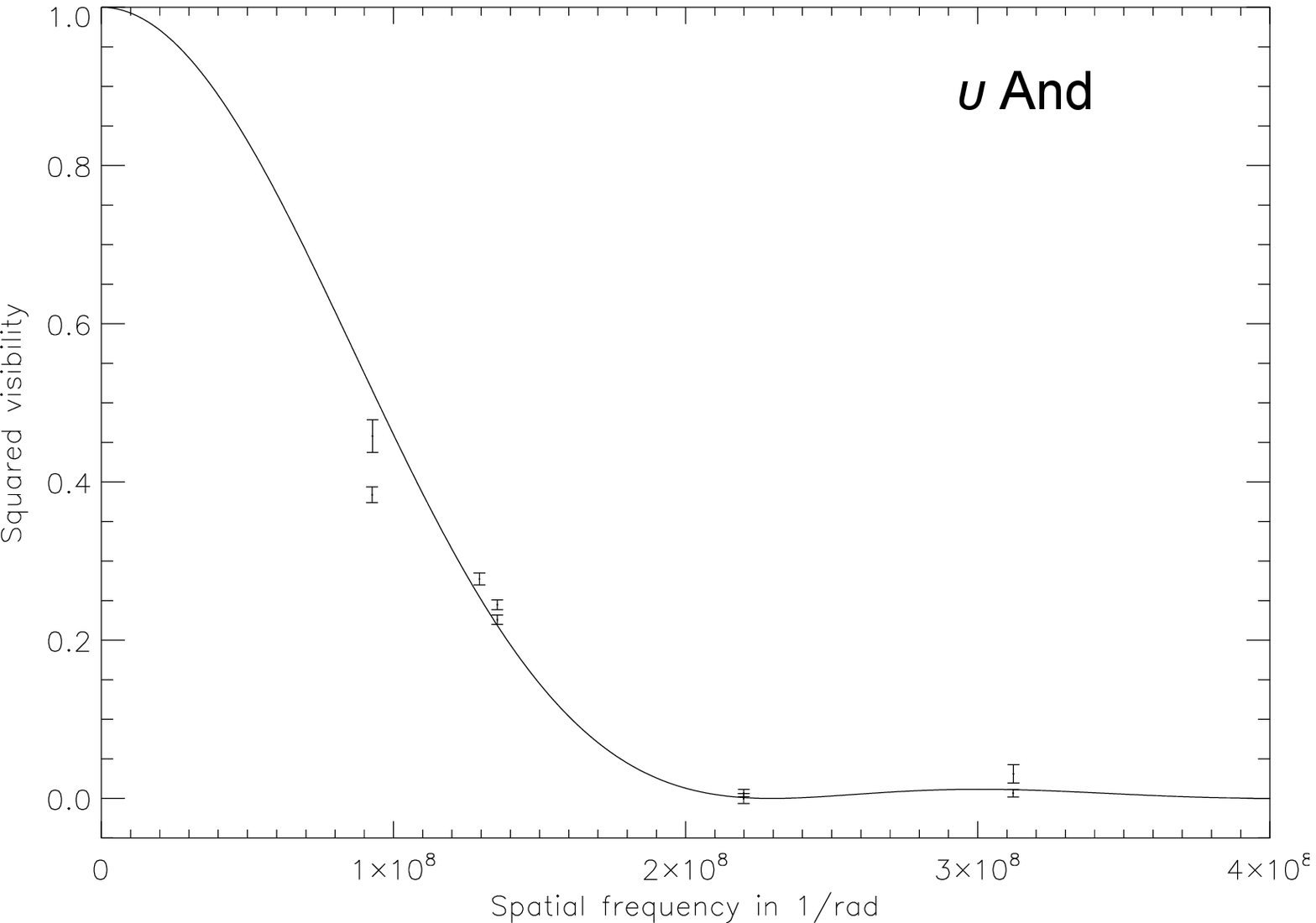}
	\includegraphics[width=8.5 cm]{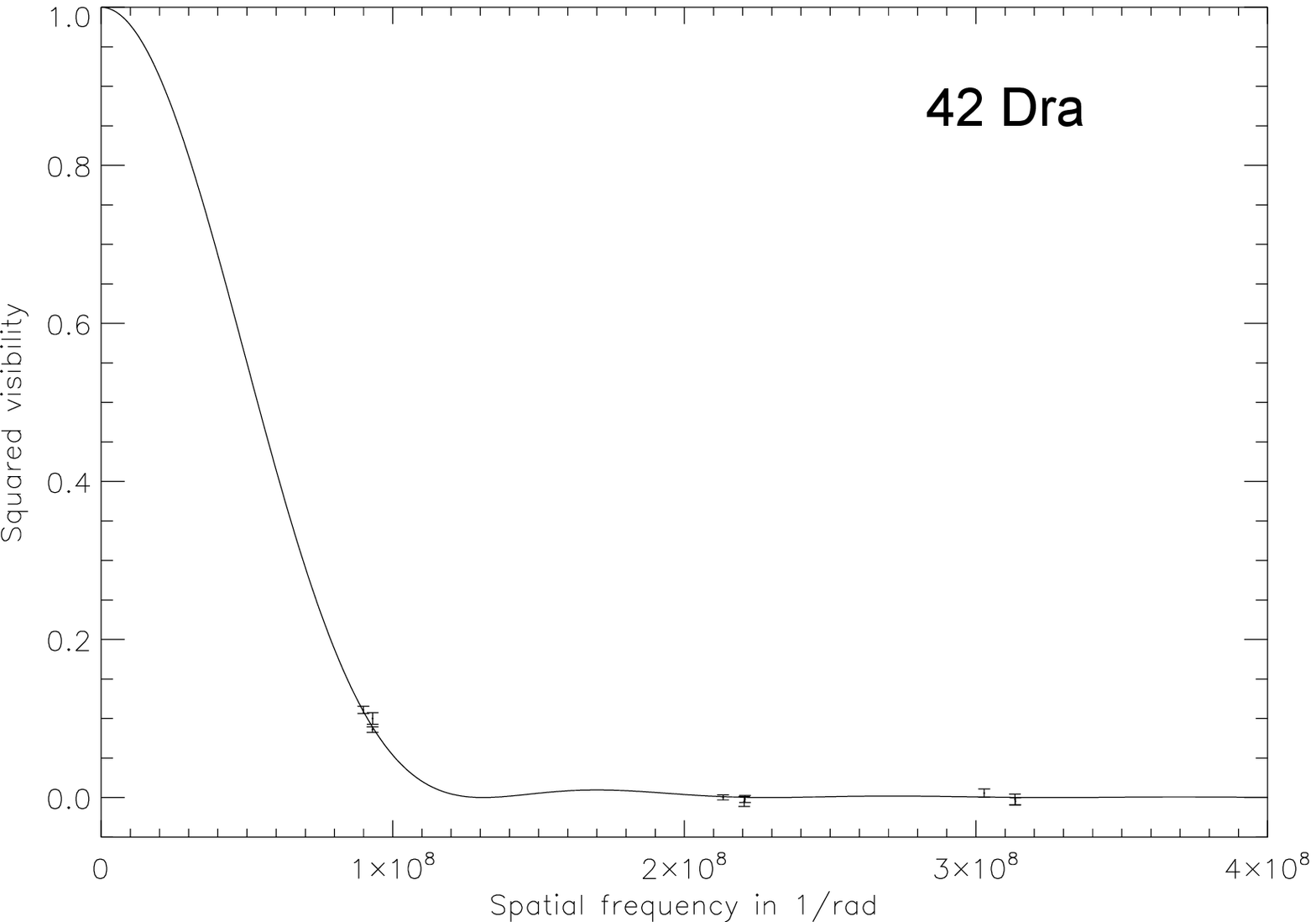}
	\caption{Squared visibility of 14~And (top), $\upsilon$~And (middle) and 42~Dra (bottom) $\textit{versus}$ spatial frequency [1/rad] for VEGA data points. The solid line is the model of the limb-darkened angular diameter provided by the LITpro software.}
\label{fig:HostStars}
\end{figure}

\begin{itemize}
\item \textit{$14$ And} \\
This star is well-fitted by a limb-darkened diameter model that provides a $\chi^2_{\rm reduced}$ of $2.8$ (see Figure~\ref{fig:HostStars}). It is obtained with the Claret coefficient $u_{\lambda}=0.700$, defined by the effective temperature and the $\log(g)$ given in Table~\ref{tab:table5}. It follows a LDD of  $1.51\pm0.02$~mas. 
\cite{baines} found an LDD of $1.34\pm0.01$~mas for 14~And, which is smaller by $\sim 10\%$ than the one we found with VEGA. But we recorded the data in the V band, whereas their values were recorded in the K band. 
\cite{14And} found that 14~And's exoplanet minimum mass is $M_{\rm pl} \sin(i) = 4.8 M_{\rm Jup}$, which is close to our result (see Table~\ref{tab:table10}), but was derived from radial velocity data, which induces a different bias.
\end{itemize}

\begin{itemize}
\item \textit{$\upsilon$ And} \\
The data points obtained at low spatial frequency are slightly lower than the LDD model. This explains the higher $\chi^2_{\rm reduced}$ than for the other stars, which equals $6.9$ (Figure~\ref{fig:HostStars}). Then, we obtained $\theta_{\rm LD}=1.18\pm0.01$~mas using $u_{\lambda}=0.534$.
$\upsilon$~And was observed by \cite{vanBelleAndvonBraun} with the Palomar Testbed Interferometer (PTI), who estimated its LDD to be $1.02\pm0.06$~mas. \cite{Baines2008} found a higher diameter with CHARA/CLASSIC \citep{Classic}\,: $1.11\pm0.01$~mas. However, it appears that, due to the dispersion in their measurements, the value of their error bars could be underestimated. In our case, the formal uncertainty is also very small but the high value of the $\chi^2_{\rm reduced}$ indicates a poor adjustment by this simple model.
No value is consistent with the respective other, ours being separated from \cite{Baines2008}'s by more than $5 \sigma$. More observations are definitively necessary to improve the accuracy and reliability of these measurements.

However, the minimum masses of $\upsilon$~And's exoplanets are consistent with those calculated by \cite{Curiel} and \cite{Wright}, but remain lower by $\simeq 10\%$ on average, when we use the orbital periods, semi-amplitudes, and eccentricities they both give (Table~\ref{tab:table10}). 
\end{itemize}

\begin{itemize}
\item \textit{$42$ Dra} \\
The $\chi^2_{\rm reduced}$ we obtained for 42 Dra is our lowest\,: $0.2$. The LDD model perfectly fits the data points. This leads to a $\theta_{\rm LD}$ of $2.12\pm0.02$~mas with a Claret coefficient of $u_{\lambda}=0.725$.
\cite{Baines2010} found a similar LDD to ours for 42~Dra\,: $2.04\pm0.04$~mas. Given the few studies of this stars, this additional measurement brings a new accurate confirmation of the diameter. Concerning the planet's fundamental parameter, we found a similar $M_{\rm pl} \sin(i)$ to that calculated by \cite{42Dra} (see Table~\ref{tab:table10}). 
\end{itemize}

Because CLIMB works in the K band, we used the corresponding Claret coefficients to estimate the LDD in this spectral band, resulting in $u_{\lambda}=0.321$, $u_{\lambda}=0.247$ and $u_{\lambda}=0.353$ for 14~And, $\upsilon$ And and 42~Dra, respectively.
In each case we used the effective temperature and the $\log(g)$ given in Table~\ref{tab:table5}.
Because CLIMB data are not very sensitive to limb darkening, because of the relatively low spatial frequencies and the fact that there is less limb darkening in K band, we used the visible coefficient for the global (VEGA+CLIMB) analysis. Although the $\chi^2_{\rm reduced}$ becomes slightly lower when including CLIMB data (Table~\ref{tab:VegaClimb}), the final results for the LDD are not changed, as expected because of the lower precision of the CLIMB visibilities and the lower influence on the diameter of the low spatial frequencies. The global results (VEGA+CLIMB) are thus the same as those obtained with VEGA only. The CLIMB data did not bring any improvements for this study.

\begin{table}
\caption{Summary of the 14~And, $\upsilon$~And and 42~Dra limb-darkened diameters (mas) calculated for VEGA data, CLIMB data and both instruments.\\
}
	\label{tab:VegaClimb}
\setlength{\tabcolsep}{4pt}
	\begin{tabular}{lllllll}
\hline\hline
\centering

 & \multicolumn{2}{c}{VEGA} & \multicolumn{2}{c}{CLIMB} & \multicolumn{2}{c}{VEGA+CLIMB} \\
Star & $\theta_{\rm LD}$ & $\chi^2$ & $\theta_{\rm LD}$ & $\chi^2$ & $\theta_{\rm LD}$ & $\chi^2$ \\

\hline
 14~And & 1.51$\pm$0.02 & 2.8 & 1.30$\pm$0.13 & 1.5 & 1.50$\pm$0.02 & 2.1 \\
 $\upsilon$ And & 1.18$\pm$0.01 & 6.9 & 0.96$\pm$0.16 & 0.8 & 1.17$\pm$0.01 & 4.6 \\
 42~Dra & 2.12$\pm$0.02 & 0.2 & 2.10$\pm$0.27 & 0.4 & 2.12$\pm$0.02 & 0.3 \\
\hline
	\end{tabular}
\end{table}

\begin{table}
\caption{Summary of the fundamental parameters of 14~And, $\upsilon$~And and 42~Dra calculated using VEGA interferometric data. $\theta_{\rm LD}$ is the limb-darkened diameter in mas. The radius and mass are given in solar units and  $T_{\rm eff}$ is given in K.\\
}
	\label{tab:table9}
	\begin{tabular}{llllll}
\hline\hline
\centering
Star & Radius & Mass & $\mathrm{T_{eff}}$ \\
\hline
 14~And & 12.82$\pm$0.32 & 2.60$\pm$0.42 & 4450$\pm$78 \\
 $\upsilon$ And & 1.70$\pm$0.02 & 1.12$\pm$0.25 & 5819$\pm$78 \\
 42~Dra & 22.04$\pm$0.48 & 0.92$\pm$0.11 & 4301$\pm$71 \\
\hline
	\end{tabular}
\end{table}

\begin{table*}
\begin{center}

\caption{Calculated exoplanets masses of 14~And, $\upsilon$ And and 42~Dra from interferometric data and comparison with previous work 
($^{(a)}$\citealt{14And}\,; $^{(b)}$\citealt{Curiel}\,; $^{(c)}$\citealt{42Dra}).}
\label{tab:table10}

\begin{tabular}{llllll}
\hline\hline
\centering

Planet & $P_{\rm orb} [days]$ & K $[m.s^{-1}]$ & e & \multicolumn{2}{c}{$M_{\rm pl} \sin(i)  [M_{\rm Jup}]$} \\
   & & & & This work & Previous work \\

\hline
14~And~b & 185.84$\pm$0.23 & 100.0$\pm$1.3 & 0 & 5.33$\pm$0.57 & 4.8$^{(a)}$ \\
$\upsilon$ And b & 4.62$\pm$0.23 & 70.51$\pm$0.45 & 0.022$\pm$0.007 & 0.62$\pm$0.09 & 0.69$\pm$0.04$^{(b)}$ \\
$\upsilon$ And c & 241.26$\pm$0.64 & 56.26$\pm$0.52 & 0.260$\pm$0.079 & 1.80$\pm$0.26 & 1.98$\pm$0.19$^{(b)}$ \\
$\upsilon$ And d & 1276.46$\pm$0.57 & 68.14$\pm$0.45 & 0.299$\pm$0.072 & 3.75$\pm$0.54 & 4.13$\pm$0.29$^{(b)}$ \\
$\upsilon$ And e & 3848.86$\pm$0.74 & 11.54$\pm$0.31 & 0.0055$\pm$0.0004 & 0.96$\pm$0.14 & 1.06$\pm$0.28$^{(b)}$ \\
42~Dra~b & 479.1$\pm$6.2 & 112.5 & 0 & 3.79$\pm$0.29 & 3.88$\pm$0.85$^{(c)}$ \\
\hline
\end{tabular}

\end{center}
\end{table*}

\section{Interferometric observations of $\theta$~Cygni with VEGA/CHARA}
\label{sect:observations}

\subsection{$\theta$~Cygni}

$\theta$~Cyg (HD185395, $d=18.33\pm0.05$ pc, Table~\ref{tab:table1}) is an F4V star with an M-dwarf companion of $0.35~M_{\odot}$ orbiting at a projected separation of $2''$ ($\simeq~46$ AU) and with a differential magnitude of $4.6$~mag in the H band. Using the data provided by \cite{Delfosse2000}, this $dM$ translates into $7.9$~mag in the V band \citep{13cygDesort}. More recently, \cite{maui} published adaptative optics (AO) data obtained with the AEOS telescopes in 2002, and reported a differential magnitude in the Bessel I-band of $5.89\pm0.089$ and a separation of $2.54''$.
This is compatible with a contrast of $\simeq 7$ at the V band. Spectroscopic data of $\theta$~Cyg collected with ELODIE and SOPHIE at the Observatoire de Haute-Provence (OHP) revealed quasi-periodical radial velocity variations with a period of approximately 150 days. No known stellar variation modes can explain such long-term, high-amplitude RV variations. They were tentatively attributed to the presence of more than two exoplanets, possibly interacting with each other. However, this explanation was not only unsatisfactory because it is dynamically unstable, but also because it did not straightforwardly explain a peak observed in the periodogram of the bisector velocity span at $\simeq 150$ days. Clearly, the data at hand were not sufficient to fully understand this complex system.

Our interferometric observations in the visible wavelengths have both high spatial and spectral resolution and help us probe the same domain as these spectroscopic results. Furthermore, we obtained measurements very close to the first zero of the visibility function (see Section~\ref{subsect:visibility}), which allows accurate angular diameter determination and the possible identification of stellar pulsations. As a Kepler target, photometric observations were obtained in 2010 and solar-like oscillations were detected \citep{Guzik2}. These observations imply the possible presence of $\gamma$ Dor gravity modes, which are generally present in early-F spectral type stars. If these oscillations are confirmed, $\theta$~Cyg would be the first star to show signs of both solar-like and $\gamma$ Dor oscillations \citep[][and references therein]{guzik}.

\begin{table}
	\caption{$\theta$~Cyg, coordinates and parameters. References from $^{(a)}$\cite{13cygDesort}\,: $^{(b)}$\cite{Boyajian2012}\,: $^{(c)}$\cite{vanBelle}\,: $^{(d)}$\cite{Guzik2}\,: $^{(e)}$\cite{Erspamer}\,: $^{(f)}$\cite{leeuwen}}.
	\label{tab:table1}
    \centering
	\begin{tabular}{lll}
\hline\hline
\multicolumn{2}{l}{Coordinates} \\
\hline
RA (J2000)  & 19 : 36 : 26.5 \\
Dec (J2000) &  + $50^{\circ}$13'16''\\
\hline
\centering
Stellar parameters & \multicolumn{2}{c}{Values}\\
\hline
Stellar type & F4V \\
V mag  & 4.50$\pm$009 & \\
K mag  & 3.5$\pm$0.296 & \\
$M_{V}$ & 3.14 \\
$v\,\sin i$ [km/s] & 7 \\
$T_{\rm eff}$ [K] & 6745 $^{(a)}$ & 6381$\pm$65 $^{(b)}$ \\
Distance [pc] & 18.33$\pm$0.05 & \\
Parallax [mas] & 54.54$\pm$0.15$^{(f)}$ \\
Radius [$R_\odot$] & 1.70$\pm$0.03 $^{(b)}$ &  \\
Mass [$M_\odot$] & 1.38$\pm$0.05 $^{(a)}$ & 1.34$\pm$0.01 $^{(b)}$ \\
Age [Gyr] & 1.5 $\pm^{+0.6}_{-0.7}$ $^{(a)}$ & 2.8$\pm$0.2 $^{(b)}$ \\
log g & 4.2 $^{(e)}$ & \\
$[Fe/H]$ & $-$0.08 $^{(a)}$ & $-$0.04 $^{(b)}$ \\
log L [$L_\odot$] & 0.63$\pm$0.003 $^{(d)}$ & 4.265$\pm$0.090$^{(b)}$ \\
\hline
\end{tabular}
\end{table}

\subsection{VEGA observations}

We performed nine observations of $\theta$~Cyg with VEGA from June 2010 to October 2011. We used the three-telescope capabilities of the instrument \citep{Vega2}, using the telescope combinations E1E2W2, W1W2E2 and W1W2E1 triplets of the CHARA array.

Three stars were used as calibrators\,: HD177003 (\textit{cal1}), HD177196 (\textit{cal2}) and HD203245 (\textit{cal3}), whose parameters are summarized in Table~\ref{tab:table2}.

	\begin{table}[h]
	\caption{Calibrators used for $\theta$~Cyg observations. The value of the equivalent uniform disk $\theta_{\rm UD}$ is given at 700nm.}
	\label{tab:table2}
	\begin{tabular}{llllll}
\hline\hline
$\sharp$ & Name & Spectral Type & $m_{V}$ & $m_{K}$ & $\theta_{\rm UD} [mas]$\\
\hline
	1 & HD 177003 & B2.5IV & 5.37 & 5.89 & 0.13$\pm$0.01 \\
	2 & HD 177196 & A7V & 5.01 & 4.51 & 0.43$\pm$0.03 \\
	3 & HD 203245 & B6V & 5.74 & 6.10 & 0.14$\pm$0.01 \\
\hline
	\end{tabular}
	\end{table}

If the target was observed only once during a night, the observing sequence was \textit{Cal1-Target-Cal1}, each calibrator observation being 20 to 40 blocks of 1000 frames long, depending on the magnitude, that is between about 10 and 20 minutes, while each target observation was 60 blocks of 1000 frames long, or about 30 minutes. When the target was observed twice, the observing sequence was either \textit{Cal1-Target-Cal2-Target-Cal2}, or \textit{Cal2-Cal1-Target-Cal1-Cal2-Target-Cal3}.
The data were recorded at medium spectral resolution and the data processing was performed on 15 to 30 nm wide channels in the continuum. The calibrated visibilities are presented in Table~\ref{tab:table3}. To take into account the variation of the spatial frequency due to the width of the spectral band (bandwith smearing effect), we calculated its effect on the visibility. We found it to be totally negligible \citep[variation lower than the error bars of the measurements,][]{Vega1}. Moreover, the data processing was performed with the same parameters for one observing sequence and effects such as these will largely calibrate out. For most of these observations, interferometric data in the infrared wavelength (K band) were also recorded with CLIMB, which was used as a 3T group delay fringe tracker \citep{climb}. However, the baselines chosen for VEGA were too small for this object to be resolved in K band and the CLIMB data were not used in the final analysis.

\begin{table}
	\caption{Journal of the observations of $\theta$~Cyg. RJD is the reduced Julian day, \textit{$\lambda_{0}$} is the central wavelength in nm, \textit{$\Delta\lambda$} is the width in nm of the analyzed spectral band. Column~4 (entitled Seq) indicates the observing and calibration strategy, with the target (T) and the associated calibrator (1, 2 or 3). The projected baseline is given by Base (in meter) and PA (in degree). $V^{2}$ is the calibrated squared visibility with a total error including statistical and systematic errors. They all represent 3-T measurements.}
	\label{tab:table3}
	\begin{tabular}{lllllll}
\hline\hline
RJD & $\lambda_{0}$ & $\Delta\lambda$ & Seq & Base & PA & $V^{2}$ \\
\hline
	55849.62 & 707.5 & 15 & 1T3 & 106 & 84.6 & 0.534 $\pm$ 0.022 \\
	  & & & & 156 & -131.9 & 0.237$\pm$0.015 \\
	  & & & & 249 & -134.5 & 0.028$\pm$0.021 \\	
	55848.62 & 707.5 & 15 & 1T3 & 106 & 83.9 & 0.502$\pm$0.020 \\
	  & & & &  156 & -132.6 & 0.192$\pm$0.007  \\
	  & & & & 249 & -135.5 & 0.048$\pm$0.014  \\	
	55826.67 & 737.0 & 14 & T1 & 66 & -139.1 & 0.801$\pm$0.038  \\
	  & & & & 156 & -132.3 & 0.229$\pm$0.016 \\
	  & & & & 221 & -134.3 & 0.054$\pm$0.028  \\
	55826.74 & 737.0 & 14 & 1T & 65 & -157.3 & 0.822$\pm$0.036  \\
	  & & & & 152 & -150.8 & 0.286$\pm$0.014  \\
	  & & & & 216 & -152.7 & 0.017$\pm$0.019  \\	
	55805.75 & 737.5 & 15 & 1T1 & 65 & -143.9 & 0.885$\pm$0.023 \\
	  & & & & 155 & -137.2 & 0.236$\pm$0.011 \\
	  & & & & 220 & -147.7 & 0.039$\pm$0.022 \\
	55803.77 & 737.5 & 15 & 3T3 & 103 & 75.6 & 0.549$\pm$0.011 \\
      & & & & 154 & -141.3 & 0.195$\pm$0.012 \\
      & & & & 245 & -146.6 & 0.040$\pm$0.018 \\	
	55774.73 & 709.5 & 15 & 1T1 & 106 & 109.9 & 0.481$\pm$0.015 \\
      & & & & 153 & -107.4 & 0.130 $\pm$0.010 \\	
	55722.93 & 735.0 & 20 & 21T12 & 108 & 95.1 & 0.451$\pm$0.015 \\
	  & & & & 156 & -121.3 & 0.166$\pm$0.009 \\
    55722.98 & 735.0 & 20 & 12T3 & 106 & 82.1 & 0.493$\pm$0.013 \\
	  & & & & 155 & -134.5 & 0.181$\pm$0.008 \\
	55486.71 & 670.0 & 20 & 1T1 & 64 & -167.3 & 0.813$\pm$0.016 \\
	  & & & & 150 & -161.0 & 0.169$\pm$0.009 \\
	  & & & & 214 & -162.9 & 0.027$\pm$0.019 \\
	55486.74 & 670.0 & 20 & 1T1 & 64 & 179.7 & 0.928$\pm$0.020 \\
	  & & & & 148 & -174.0 & 0.166$\pm$0.010 \\
	55370.92 & 715.0 & 30 & T2 & 66 & -133.8 & 0.788$\pm$0.028 \\
	  & & & & 156 & -127.0 & 0.152$\pm$0.013 \\
	  & & & & 222 & -129.0 & 0.012$\pm$0.010 \\
	55370.96 & 715.0 & 30 & 2T2 & 65 & -148.4 & 0.802$\pm$0.030 \\
	  & & & & 154 & -141.7 & 0.221$\pm$0.019 \\
	  & & & & 219 & -143.7 & 0.039$\pm$0.015 \\
	
\hline
	\end{tabular}
\end{table}

\section{Determination of $\theta$  Cygni's fundamental parameters}
\label{sect:parameters}

\subsection{Determination of the limb-darkened diameter}

\begin{figure}[h]
	\includegraphics[scale=0.50]{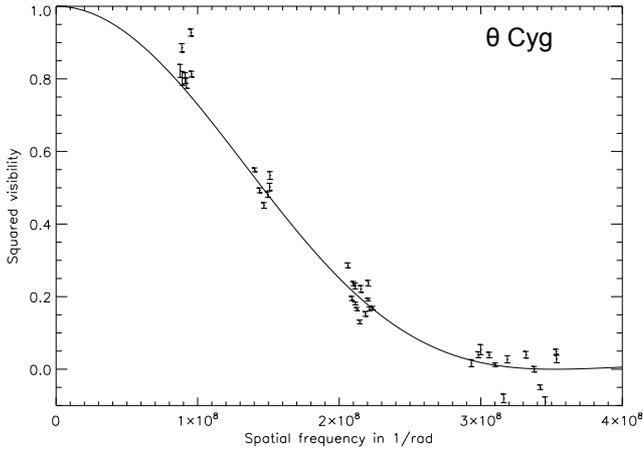}
	\caption{Squared visibility of $\theta$~Cyg as a function of the spatial frequency [1/rad] for all data points recorded in 2010 and 2011 by VEGA (3T configuration). The solid line is the squared visibility function for a linear limb-darkened disk model with a diameter of $0.76$~mas and a limb-darkening coefficient of $0.5$, obtained with LITpro software.}
	\label{fig:13Cyg}
\end{figure}

For almost all observations including the E1E2 baseline, we obtained a $\chi^2_{\rm reduced}$ larger than 2. The E2 telescope is known to present instabilities, like vibrations and delay line cart problems. Those points are therefore more dispersed than those at higher spatial frequencies, as shown in Figure~\ref{fig:13Cyg} and Table~\ref{tab:table11}, and the value of $\chi^2_{\rm reduced}$ is mostly dominated by these points.
In a first analysis, we have considered all data points. We used the LitPro software\footnote{Available at http://www.jmmc.fr/litpro} \citep{litpro} and obtained a mean UD equivalent diameter of $0.726\pm0.003$~mas.
The $\chi^2_{\rm reduced}$ of the model fitting is equal to $8.4$, which clearly indicates dispersion in the measurements or possible variations of the diameter from night to night. This will be investigated in Sect.~\ref{sect:discussion}. We also tested a linear limb-darkened (LD) disk model with a coefficient $u_{\lambda}$ as defined by \cite{Hanbury}. Unfortunately, the data quality at low-visibility levels is not sufficient for a correct $u_{\lambda}$ determination. For a more detailed analysis, we decided to fix the linear LD coefficient in the LitPro software. With $T_{\rm eff} = 6745$~K and $\log(g) = 4.2$, we used the value of the Claret coefficients \citep{claret2011} given for the R, I and J bands, and deduced by extrapolation the value at the observing wavelengths (715 and 670 nm). We found $u_{670 nm} = 0.510$ and $u_{715 nm} = 0.493$, and finally took the mean value 0.5. The adjustment of the whole data set (see Figure~\ref{fig:13Cyg}) gives the value $\theta_{\rm LD}=0.760\pm0.003$~mas, with a reduced $\chi^2_{\rm reduced}$ equal to $8.5$.\\
Our final value is consistent with the diameter estimated by \cite{vanBelle} with spectral energy distribution based on photometric observations\,: they found $\theta_{\rm LD}=0.760\pm0.021$~mas. \cite{Boyajian2012} observed this star in 2007 and 2008 with the CHARA CLASSIC beam combiner operating in the K band, and found $\theta_{\rm LD}=0.845\pm0.015$~mas and $\theta_{\rm LD}=0.861\pm0.015$~mas, which is much larger than ours. We will return to this point in Sect.\ref{sect:discussion}.

As previously stated, the CLIMB measurements have large scatter and are at much lower spatial frequencies than the VEGA data. They provide a LDD equal to $\theta_{\rm LD}=0.654\pm0.090$~mas, with a $\chi^2_{\rm reduced} = 1.86$, obtained with a Claret coefficient of $0.22$ corresponding to the K band. When combining the CLIMB and VEGA data, the diameter remains the same, except that the $\chi^2_{\rm reduced}$ decreases to $5.3$. This is because of the large error bars obtained for CLIMB data at high frequencies, which do not constrain $\theta$~Cyg's LDD at all, although they reduce the $\chi^2_{\rm reduced}$. For $\theta$~Cyg, the scatter affects all measurements.

\subsection{Determination of fundamental parameters}

The radius and the mass of $\theta$~Cyg were estimated using Equations~\ref{eq:radius} and \ref{eq:rm1}. We took $\pi=54.54\pm0.15$~mas according to \cite{leeuwen}. $\theta$~Cyg 's radius is then $R=1.503\pm0.007 R_{\odot}$. The final uncertainty is equally due to errors in the parallax and the angular diameter.
This results in a mass of $1.32\pm0.14 M_\odot$ and locates $\theta$~Cyg between the two lines representing the evolutionary tracks of Figure 4 in the model of \cite{Guzik2}.
Finally, the effective temperature was calculated using Equation~\ref{eq:Teff} and the luminosities shown in Table~\ref{tab:table1}. The errors were calculated using the Monte Carlo method.
This results in  $T_{\rm eff} = 6767 \pm 87$~K, which is also consistent with the value given by \cite{13cygDesort}. \cite{Boyajian2012} found a lower $T_{\rm eff}$ of $6381 \pm 65$~K mostly due to a larger limb-darkened diameter (see Table~\ref{tab:table1}).
Table~\ref{tab:table4} summarizes the results based on our interferometric measurements.

\begin{table}
\caption{Table summarizing $\theta$~Cyg's fundamental parameters calculated with the interferometric data.}
	\label{tab:table4}
	\begin{tabular}{l l}
\hline\hline
\centering
Stellar parameters & Value$\pm$Error \\
\hline
LD diameter [$mas$] & 0.760$\pm$0.003 \\
Radius [$R_\odot$] & 1.503$\pm$0.007 \\
Mass [$M_\odot$] & 1.32$\pm$0.14 \\
$T_{\it{eff}}$ [$K$] & 6767$\pm$87 \\
\hline
	\end{tabular}
\end{table}

\section{Discussion}
\label{sect:discussion}

\begin{figure}[h]
	\includegraphics[scale=0.50]{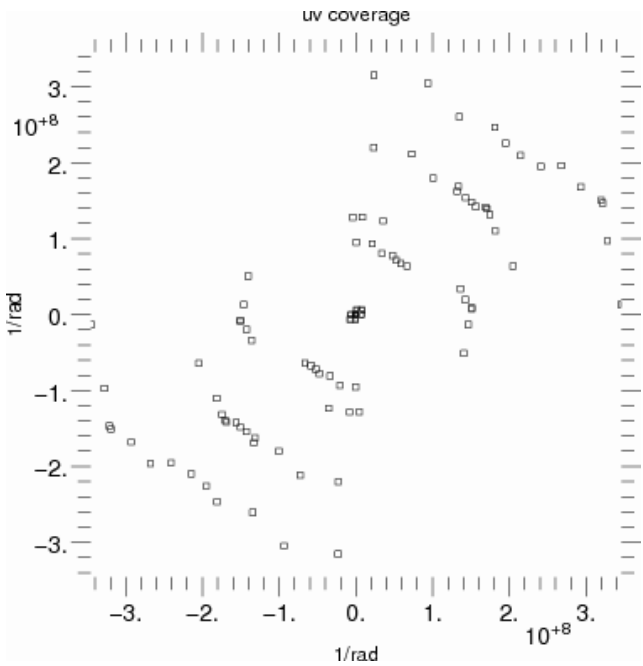}
	\caption{UV coverage of the baselines used during $\theta$~Cyg observations from 2010 to 2011.}
	\label{fig:13CygUVcoverage}
\end{figure}

We have seen in the previous section that the scatter of measurements for $\theta$~Cyg is larger than for the three other targets. It remains then to understand these variations. Table~\ref{tab:table11} shows the night-to-night variations in the LDD of $\theta$~Cyg. The UD and LD models do not fit these results very well, as indicated by the generally high value of $\chi^2_{\rm reduced}$. \cite{Boyajian2012}'s CLASSIC data obtained between 2007 and 2008 also show some discrepancies in their visibility curve fitted with a UD model. This introduces the possibility of either an additional companion, or stellar variations around $\theta$~Cyg.
The night-by-night observing strategy we employed so far was not optimized for the investigation of binarity but for the measurement of fundamental parameters. Thus, the UV coverage (Figure~\ref{fig:13CygUVcoverage}), which represents the support of the spatial frequencies measured by the interferometer, does not constrain on the position of an hypothetical companion very well.

\begin{table}
\caption{Values of the mean $\theta_{\rm LD}$ per night for $\theta$~Cyg and the corresponding $\chi{^2}_{\rm reduced}$.}
	\label{tab:table11}
	\begin{tabular}{l l l l l}
\hline\hline
Epoch & Baselines & $\theta_{\rm LD}$ & $\phi$ $\mathrm{mod 150}$ & $\chi^2_{reduced}$ \\ 
\hline
	55849.62 & W2W1E2 & 0.700$\pm$0.011 & 0.33 & 0.700 \\ 
	55848.62 & W2W1E2 & 0.744$\pm$0.007 & 0.32 & 5.698 \\ 
	55826.67 & E2E1W2 & 0.721$\pm$0.009 & 0.18 & 1.12 \\ 
	55805.75 & E2E1W2 & 0.727$\pm$0.010 & 0.04 & 7.749 \\ 
	55803.77 & W2W1E2 & 0.759$\pm$0.008 & 0.03 & 5.936 \\ 
	55774.73 & W2W1E2 & 0.807$\pm$0.010 & 0.83 & 13.9 \\ 
	55722.93 & W2W1E2 & 0.793$\pm$0.006 & 0.49 & 0.664 \\ 
	55486.71 & E2E1W2 & 0.744$\pm$0.007 & 0.91 & 23.2 \\  
	55370.92 & E2E1W2 & 0.764$\pm$0.010 & 0.14 & 2.468 \\ 
\hline
	\end{tabular}
\end{table}

\subsection{Stellar variations}

\begin{figure}[h]
	\includegraphics[scale=0.50]{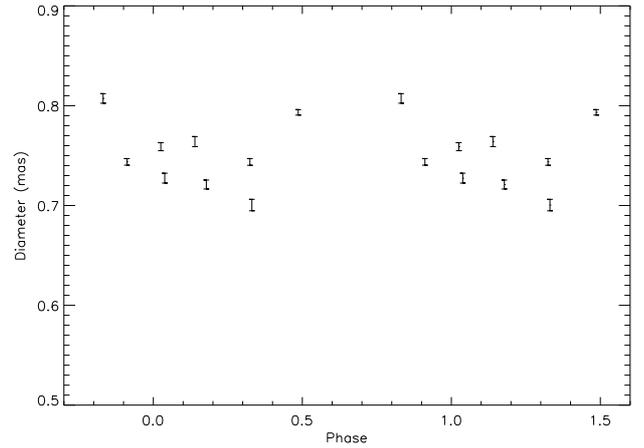}
	\caption{Individual angular diameter measurements of $\theta$~Cyg according to the phase. The phase is proportional to the reduced Julian day \textit{modulo} 150, as the radial-velocity period is expected to be.}
\label{fig:variation}
\end{figure}

Because $\theta$~Cyg's radial velocity is suspected to have a 150-day period \citep{13cygDesort}, we studied a possible correlation between the variation of the diameter and this periodic behavior of the radial velocities. Figure~\ref{fig:variation} represents the individual angular diameter plotted as a function of a phase ($\phi$) corresponding to the reduced Julian day \textit{modulo} the spectroscopic period of 150 days. This figure highlights a possible variation with an amplitude of $\sim13\%$ in diameter peak to peak. Solar-like oscillations lead to lower variations in amplitude than that, but Cepheid stars show similar-sized pulsations. According to $\theta$~Cyg's luminosity, however, it is not bright enough to be classified as a Cepheid. Moreover, a Cepheid's light curve presents much larger amplitude variations than $\theta$~Cyg's \citep[Figure 1 in][]{Guzik2}. Its luminosity and temperature would instead locate it near the instability branch of the HR diagram, identifying it as a $\delta$ Scuti or $\gamma$ Dor star, which are also A- or F- type stars.

This last possibility is also mentioned by \cite{Guzik2}, who proposed two different models that could show evidence for $\gamma$ Dor pulsations, but they only allowed $l=1$ or $l=2$ unstable g-modes. Their light curve does not reveal the typical $\gamma$ Dor frequencies around $11~\mu$Hz, which are specific for these pulsations, though they do mention that these could be overshadowed by the granulation noise. Moreover, $\gamma$ Dor oscillations have been found in many Kepler sources without much ambiguity because they show obvious evidence for this type of pulsation \citep{tkachenko}. Also, the RV measurements published in \cite{13cygDesort} do not reveal high-amplitude and high-frequency (hours to days) RV variations typical of $\gamma$ Dor stars or $\delta$ Scuti stars.

Finally, we note that if the 150-day period RV variations were due to diameter variations, these diameter variations would be unrealistically large, much larger than those observed, and very significant photometric variations should have been detected by Kepler.

We therefore conclude that stellar variations do not explain the observed features in a satisfactory manner. We therefore consider the possibility of an unseen stellar companion for $\theta$~Cyg, and see how the present interferometric data can help to test such a scenario.

\subsection{An additional companion?}

The known M-type companion to $\theta$~Cyg clearly does not affect our visibilities, because of the large separation in position (2 seconds of arc) and the large difference in magnitude (around ~7). Any instrument hosted by the CHARA array and used in the same conditions as we did (e.g., medium resolution for VEGA) has an interferometric field of view much smaller than the telescopes' Airy spot, i.e. $\simeq 0.1$ second of arc. This means that any object located beyond this field does not interfere, but could create a photometric background that disrupts the visibility of the target if it is located in the entrance field of the instrument. In our case, the $dM$ in the $V$ band gives a very small contribution to this background, much lower than the error bars ($\simeq 1\%$).
We therefore consider  the presence of a second and much closer companion. The lower limit of detection allowed by adaptive optics (AO) is at about the diffraction limit of PUEO on the CFHT, i.e. around $100$~mas for low-contrast binaries. Accordingly, a companion whose position is closer in than this limit would not be seen in AO direct imaging. However, given our current accuracies in visibility measurements, it could be detected by interferometric instruments if its flux contribution is higher than $2\%$. Because $\theta$~Cyg is not classified as SB2, such a flux ratio would imply a pole-on bound system or a visual unbound binary.

This last possibility has been considered, but is difficult to confirm. No objects are located close to $\theta$~Cyg in the background, except for $\theta$~Cyg-B, which could have moved closer to the main star over the years. As said by \cite{13cygDesort}, the differential magnitude between the two bound stars in the V band is $7.9$ mag and $4.6$ mag in the H band. Thus, we can expect a $dM$ of $\sim 3$ mag in the K band, which would make it observable with CLASSIC. A rough estimate of the orbit of $\theta$~Cyg-B based on the data published by \cite{13cygDesort} shows that at the epoch of the interferometric observations, the separation is still larger than about $2$ seconds of arc, which is well outside our interferometric field of view.

To explore the possibility of an unknown close companion around  $\theta$~Cyg, we performed several tests on our data set. Because the VEGA visibilities are, at first approximation, dominated by one main resolved source, that is the primary component, we adopted a diameter of the companion of $0.2$ mas, corresponding to an unresolved source. The UD diameter of the primary was fixed to $\theta_{\rm UD} = 0.726$ mas, which is the diameter obtained when merging all nights. Then, by assuming a companion's flux in the range 2$\%$ to 15$\%$, we obtained the position angle (PA) and angular separation ($\rho$) corresponding to the minimum $\chi^2_{\rm reduced}$. We performed the same tests with \cite{Boyajian2012}'s CLASSIC data from 2007-2008 (Table~\ref{tab:table12}).

\begin{table}
\caption{Comparison between a UD model and a model with a companion for VEGA and CLASSIC data. For each set of simulation, this table gives the orbital parameters obtained with the minimum $\chi^2_{\rm reduced}$ and the corresponding flux.}
	\label{tab:table12}
	\begin{tabular}{l l l|l l l l}
\hline\hline

& \multicolumn{2}{c}{UD model} & \multicolumn{4}{c}{Binary model} \\
Epoch & $\theta_{\rm UD} $ & $\chi^2_{\rm reduced}$ & $\rho$ & $PA$ & Flux & $\chi^2_{\rm reduced}$ \\
& [mas] & & [mas] & $[^{\circ}]$ & $\%$ & \\
\hline
VEGA\\
\hline
	55849.62 & 0.670 & 0.8 & 7.1 & 72.2 & 15 & 0.9 \\ 
	55848.62 & 0.710 & 5.7 & 11.1 & 234.7 & 15 & 5.9 \\ 
	55826.67 & 0.689 & 1.1 & 50.5 & 75.2 & 15 & 1.0 \\  
	55805.75 & 0.695 & 7.6 & 10.3 & 10.0 & 15 & 11.5 \\ 
	55803.77 & 0.726 & 5.7 &  66.7 & 182.5 & 15 & 1.3 \\ 
	55722.93 & 0.758 & 0.6 & 80.8 & 85.2 & 3 & 0.07 \\ 
	55486.71 & 0.710 & 22.7 & 34.4 & 304.8 & 15 & 20.7 \\  
	55370.92 & 0.729 & 2.5 & 50.5 & 247.7 & 10 & 0.16 \\ 
\hline
CLASSIC\\
\hline
	55794.0 & 0.762 & 0.006 & 72.7 & 115.3 & 8 & 0.02 \\
	54672.0 & 0.852 & 0.6 & 55.6 & 5.0 & 10 & 0.20 \\
	54406.0 & 0.928 & 0.03 & 86.9 & 222.6 & 7 & 0.01  \\
	54301.0 & 0.827 & 1.3 & 56.6 & 3.0 & 10 & 0.28 \\
\hline
	\end{tabular}
\end{table}

\begin{figure*}
\centering
\begin{tabular}{cc}
  \includegraphics[width=6cm,height=6cm]{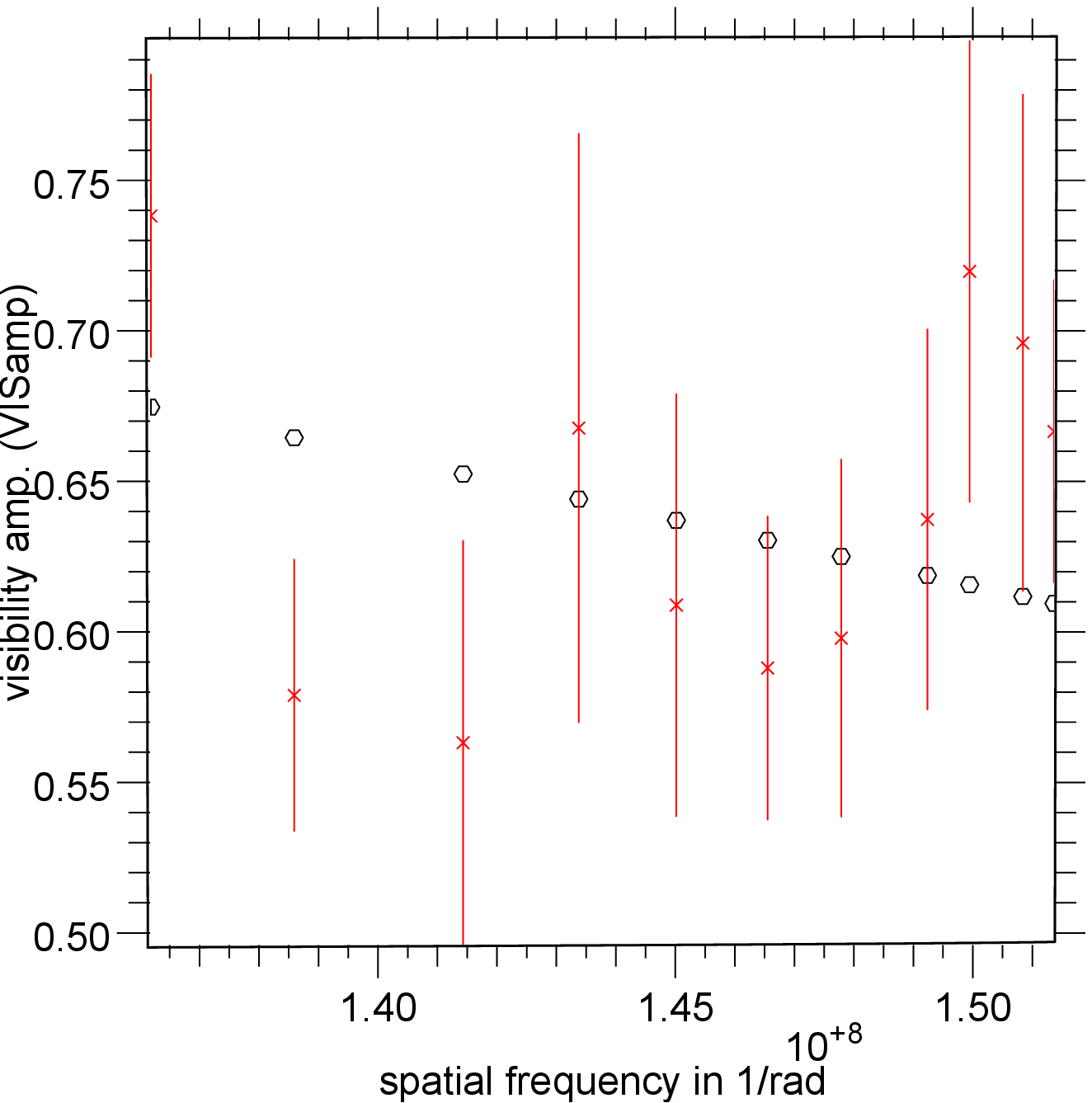}
  \includegraphics[width=6cm,height=6cm]{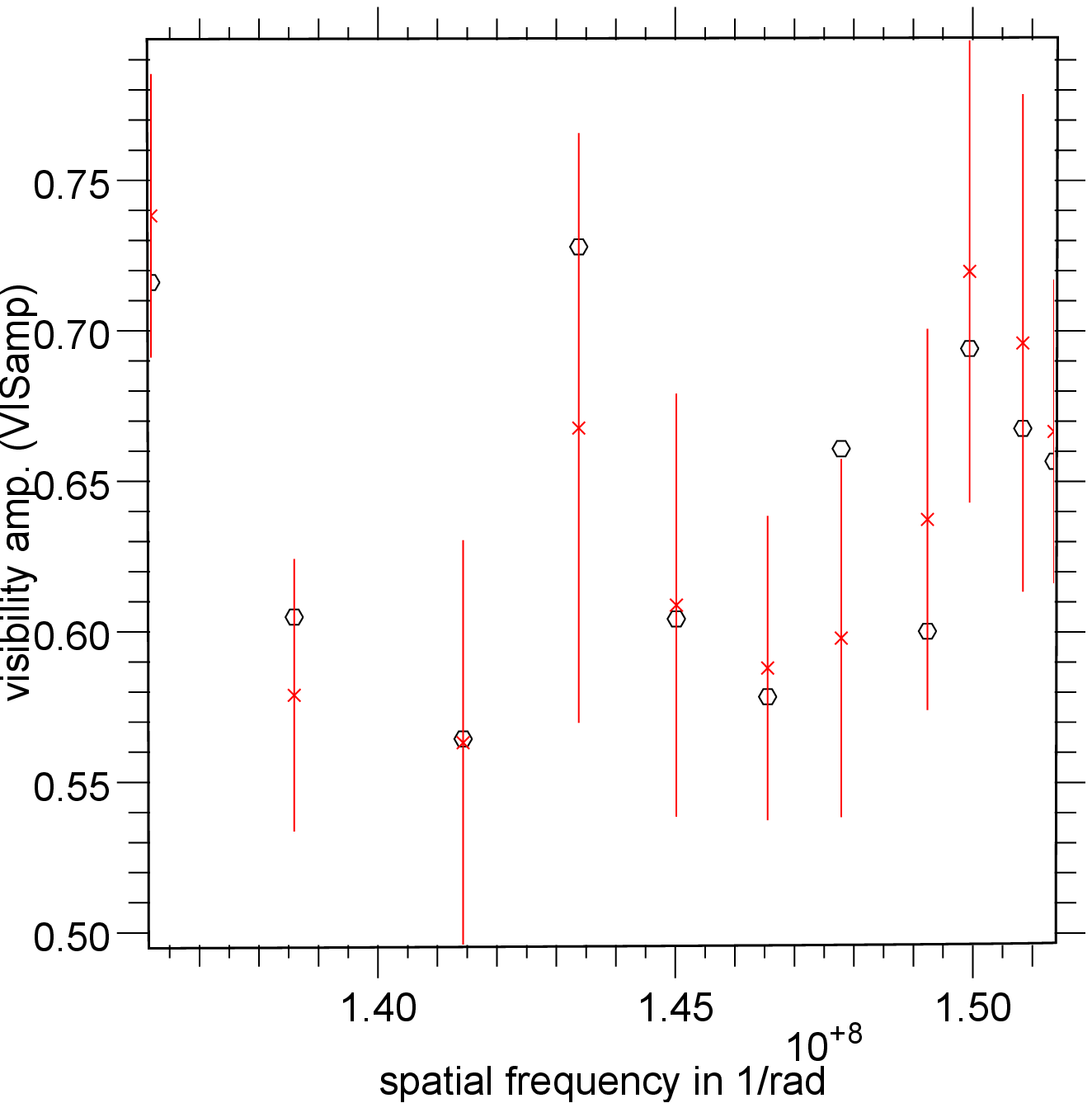}\\
  \includegraphics[width=6cm,height=4cm]{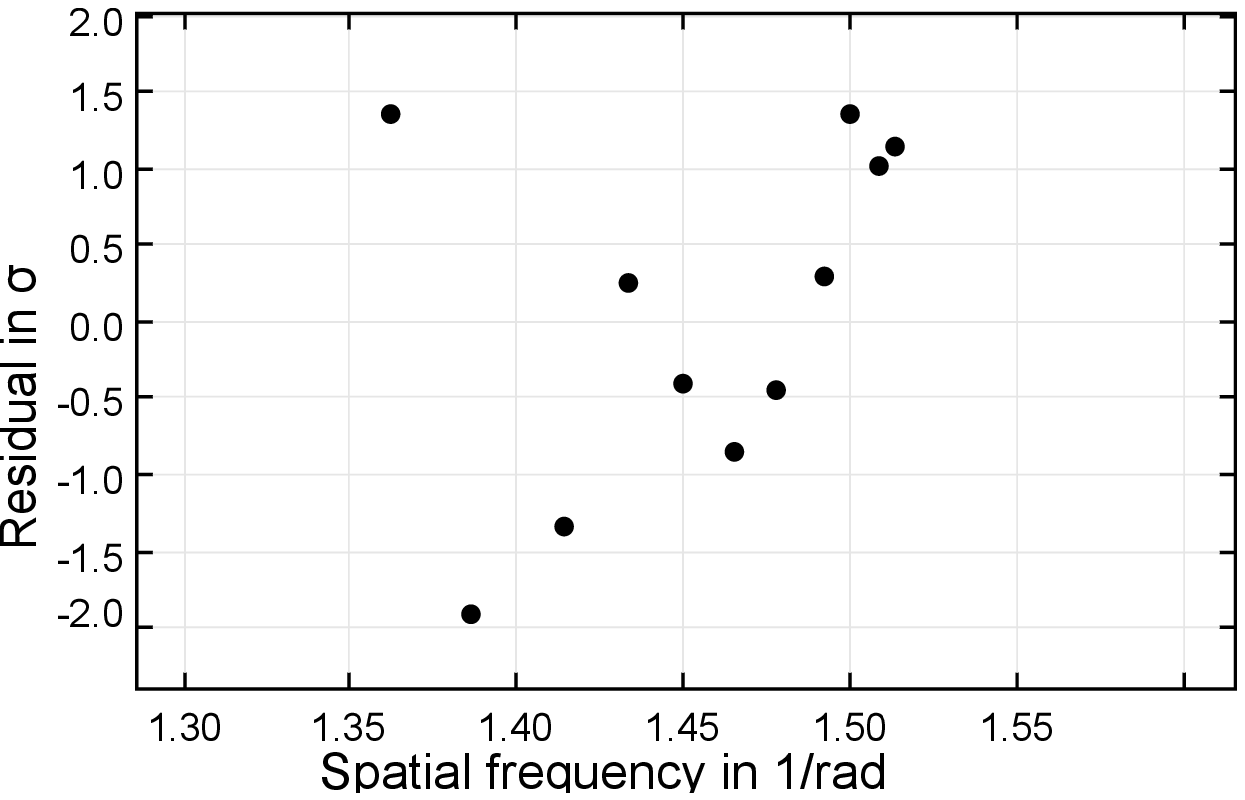}
  \includegraphics[width=6cm,height=4cm]{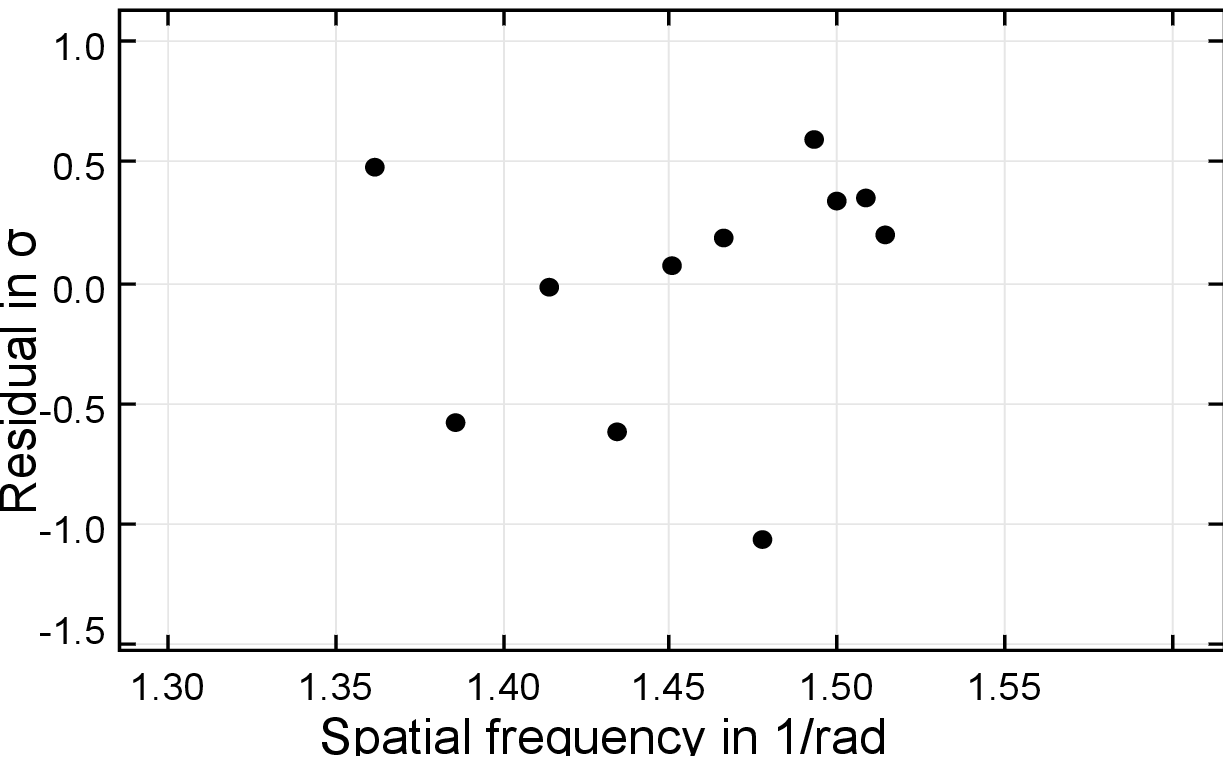}\\
\end{tabular}
  \caption{Left column\,: UD model\,: right Column\,: UD model + companion. Top row is for the CLASSIC data visibilities of the RJD 54301 (open circles for the models, crosses for the data), whereas the bottom row corresponds to the visibility residuals.}
  \label{fig:binarity}
\end{figure*}

In half of the cases of the VEGA sets, we found a solution with a better $\chi^2_{\rm reduced}$ than with a UD model. Generally, the best solution corresponds to a companion with $15\%$ of flux, and a $\rho$ included between $17.6$ and $26.9$ mas. However, in the other VEGA cases, the data do fit the binary model and no better solution is found.\\
In the CLASSIC data, the $\chi^2_{\rm reduced}$ is reduced by a factor~$2$ when we include the binarity. The better UV coverage obtained with the E1S1 baseline provides much better constraints on the model in that case. The best solution for the CLASSIC data gives a flux ratio of about $7\%$ and a separation of about $25$ mas.
An example of the fit improvement for the CLASSIC measurements is presented in Fig.~\ref{fig:binarity}. However, this flux ratio does not permit us to tell which type of star the companion could correspond to, because it is not necessarily bound, but coud be either foreground or background.

Finally, we explored the existence of a closure phase signal generated by this close unknown companion \citep{lebouquin}. The closure phase is the sum of the phases of the complex visibilities obtained with the three baselines of a triplet. It is independent of the atmosphere, giving direct information of the object's visibility, which results in informations about asymmetries, presence of a companion, etc. We already said that the CLIMB data were at low spatial frequencies due to the longer wavelength of operation. Simulations show that in the baseline configurations used for this paper, the companion will produce a signal lower than $5$ or $10^\circ$, which is below the current accuracy of CLIMB phase closure measurements. However, the simulation shows that a huge closure phase signal of $\pm 40^\circ$ should be detected by VEGA with the E1E2W2 configuration. Many tests have been performed on the data sets but the signal-to-noise ratio of VEGA phase closure measurements is not sufficient for a correct determination. Unlike from the estimation published in \cite{Vega2}, we have now a clearer understanding of the noise level in closure phase measurements with VEGA. Closure phase is a third-order moment and the multi-speckle regime of VEGA prevents us from obtaining accurate closure phase measurements for stars fainter than magnitude $1$ or $2$, depending on seeing conditions (Mourard et al., 2012 in preparation).

\section{Conclusion}

\begin{figure}
\centering
  \includegraphics[width=8cm,height=4cm]{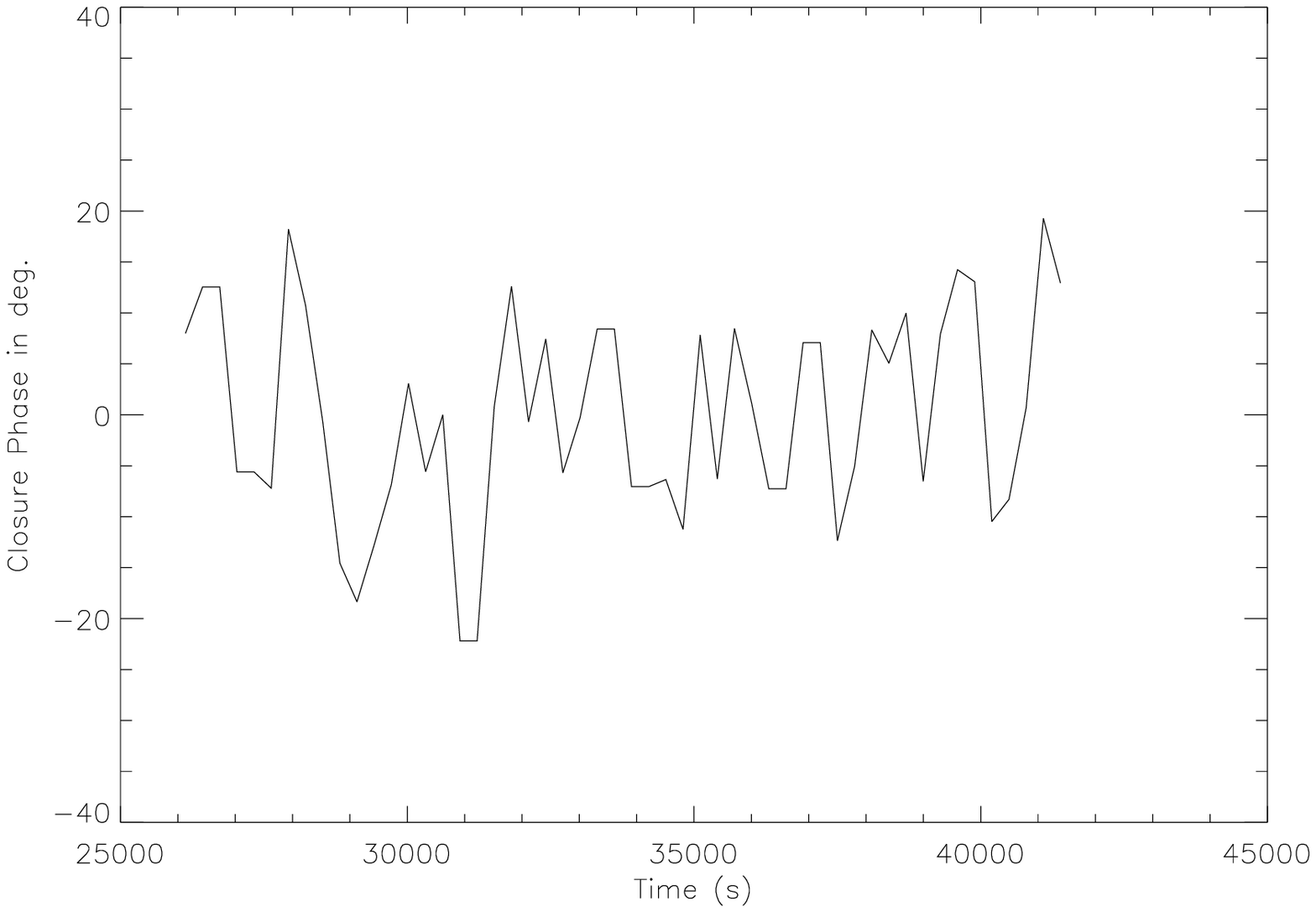}\\
  \includegraphics[width=8cm,height=4cm]{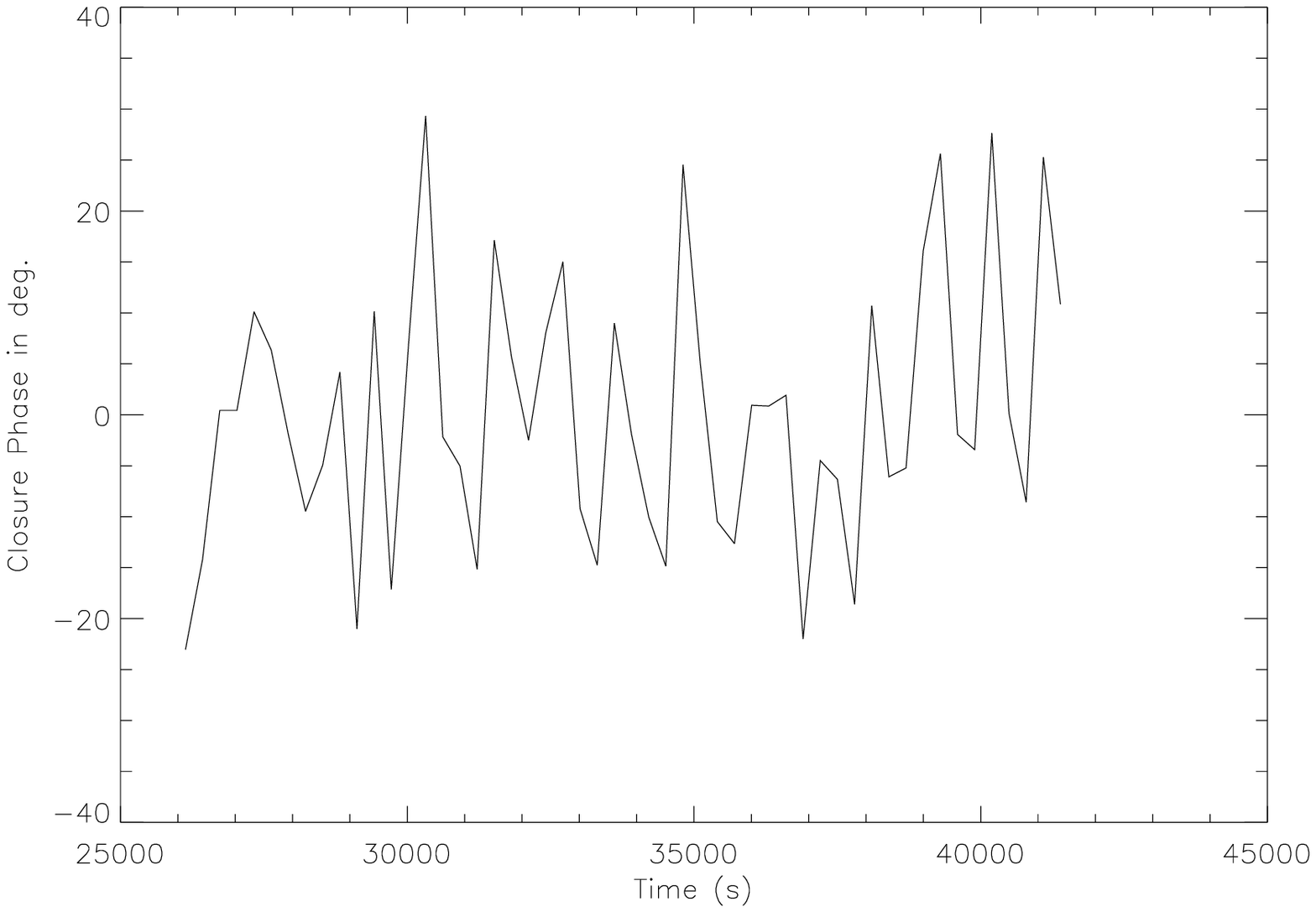}
  \caption{Closure phase signals expected for $\theta$~Cyg with a companion located at $25$ mas that contributes to $10\%$ of the flux. UP\,: CLIMB in K band, BOTTOM\,: MIRC in H band. We used the largest triangle of telescopes on CHARA\,: E1W1S1.}
  \label{fig:cloture}
\end{figure}

We obtained new and accurate visibility measurements of 14~And, $\upsilon$ And and 42~Dra using visible band interferometric observations. From these we derived accurate values of the LD diameter and of fundamental parameters that are fully consistent with those derived with other techniques and bring some improvements in precision. The error bars and $\chi^2_{\rm reduced}$ for these three stars are in general much smaller than those obtained on our fourth target\,: $\theta$~Cygni.
We analyzed the scatter of measurements of $\theta$~Cyg, taking into account that instrumental or data processing bias are well understood thanks to the good results obtained on the three other stars. It appears that a solution with an unknown companion close to the star helps in reducing the residuals in the model fitting. The limited accuracy in our determination prevents us from being conclusive about the presence of a new close companion around $\theta$~Cyg, and do not allow us to tell which type of star it could be, because it is not necessarily bound. However, this result encourages organizing new observations in the visible and IR wavelengths, focused on confirming or denying this hypothesis. Closure phase signal is a good way to detect and characterize faint companions around bright stars. We performed simulations of expected closure phase signal for $\theta$~Cyg with a companion contributing $10\%$ of the flux that is located at $25$ mas. As explained before, VEGA is unfortunately not able to measure accurate closure phase signals. CLIMB in K band and MIRC \citep{mirc} in H band are well-adapted for closure phase tests with the largest CHARA triangle (E1W1S1). Expected signals are presented in Fig.~\ref{fig:cloture}. Therefore, a more adequate observing strategy and dedicated observations will be prepared with the combination of the different CHARA beam combiners.

\begin{acknowledgements}
The CHARA Array is operated with support from the National Science Foundation through grant AST-0908253, the W. M. Keck Foundation, the NASA Exoplanet Science Institute, and from Georgia State University. RL warmly thanks all the VEGA observers that permitted the acquisition of this set of data. RL also acknowledges the PhD financial support from the Observatoire de la C\^{o}te d'Azur and the PACA region. We acknowledge the use of the electronic database from CDS, Strasbourg and electronic bibliography maintained by the NASA/ADS system.
\end{acknowledgements}

\bibliographystyle{aa}
\bibliography{13CYG}

\Online
\begin{appendix}
\section{Individual angular diameter determinations}
\label{appendix}
We present here the different individual LD angular diameter determinations for the various epochs of observation of $\theta$ Cygni. The individual LD diameters are given in Tab.~\ref{tab:table11}.

\begin{figure}[h]
    \subfloat{\includegraphics[scale=0.5]{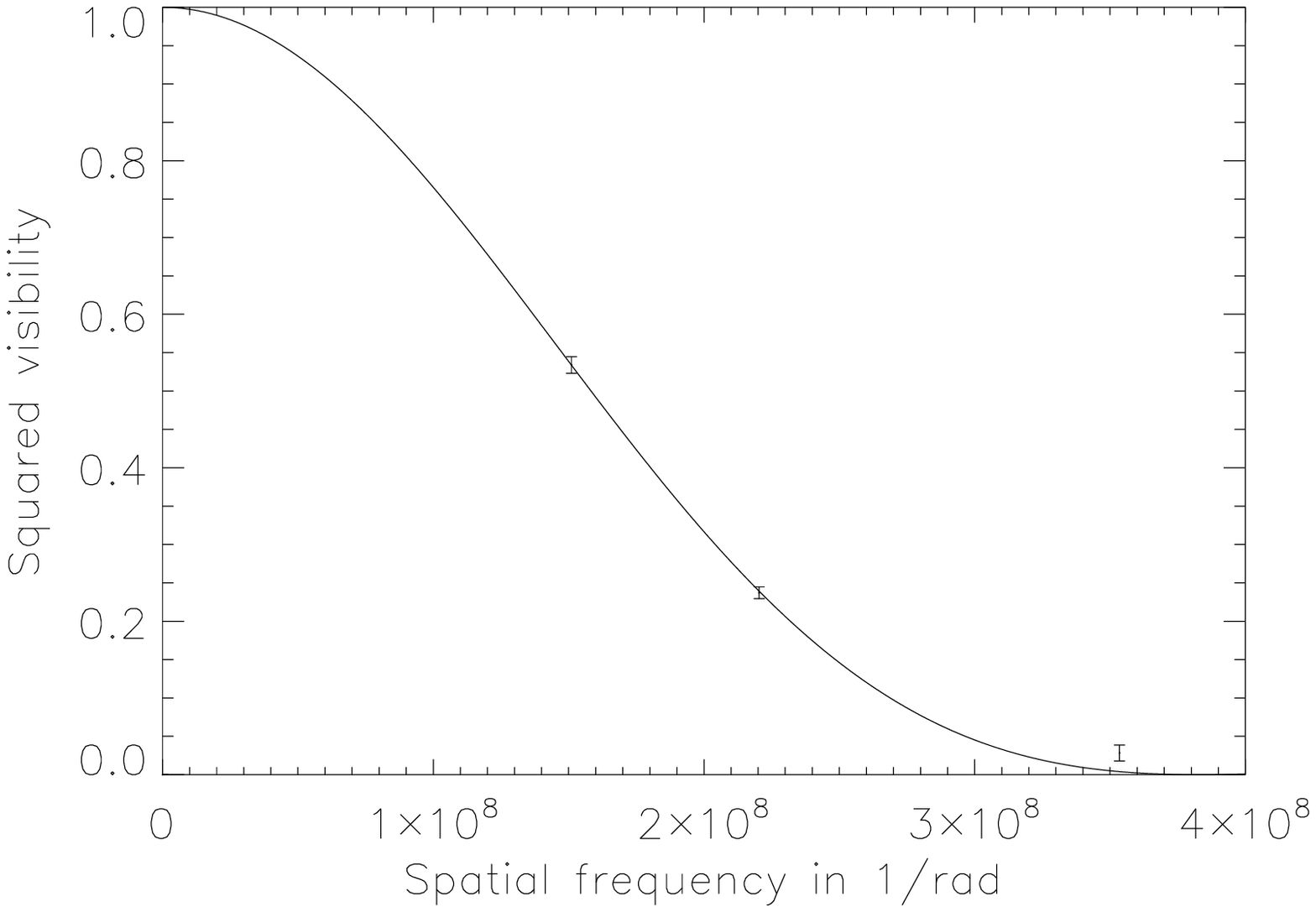}}
    \subfloat{\includegraphics[scale=0.5]{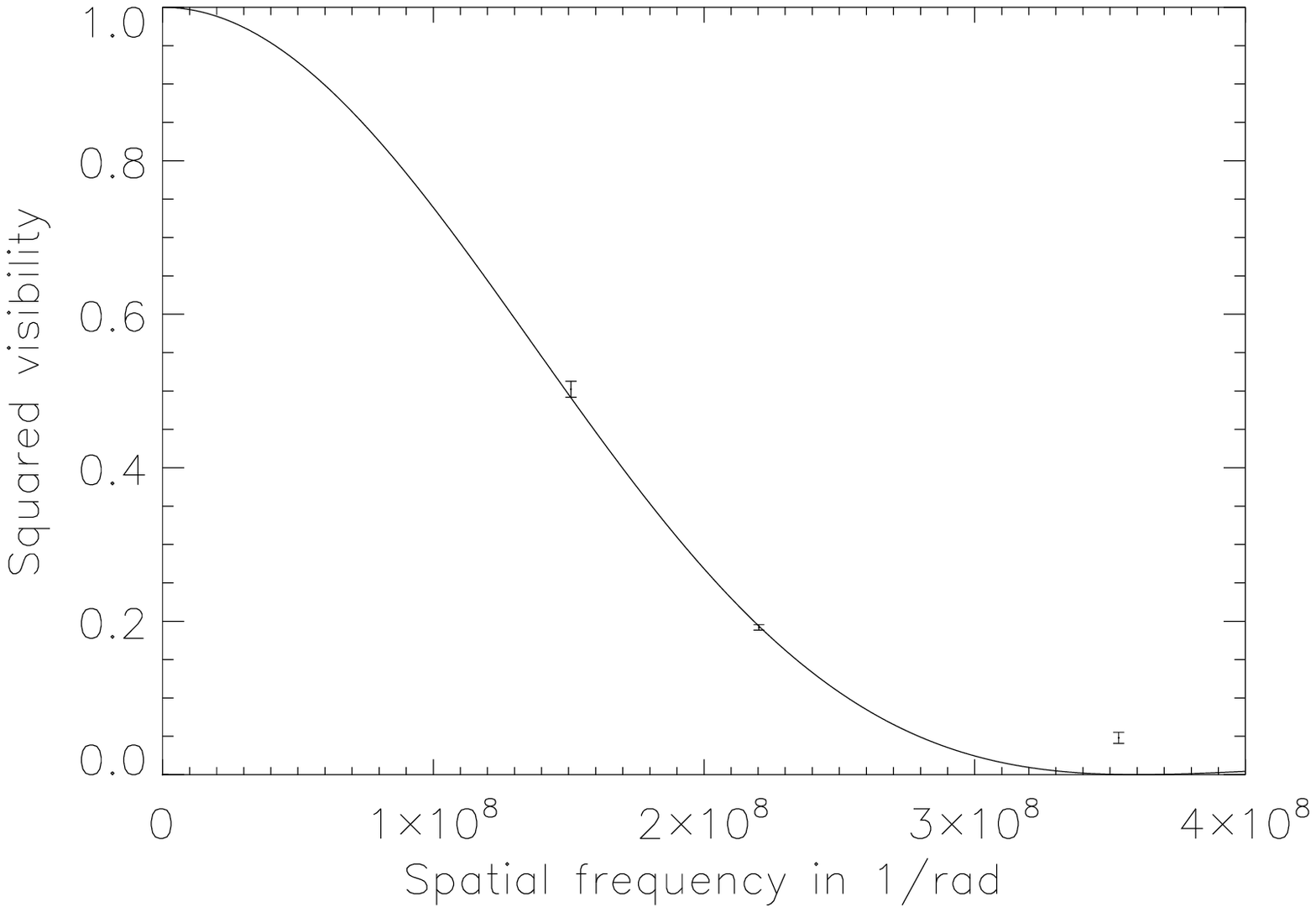}}
    \caption{Model of squared visibility for the RJD 55849.62 (left, $\theta_{\rm LD}=0.700\pm0.011$) and 55848.62 (right, $\theta_{\rm LD}=0.744\pm0.007$) obtained by LITpro.}
\end{figure}

\begin{figure}[h]
    \subfloat{\includegraphics[scale=0.5]{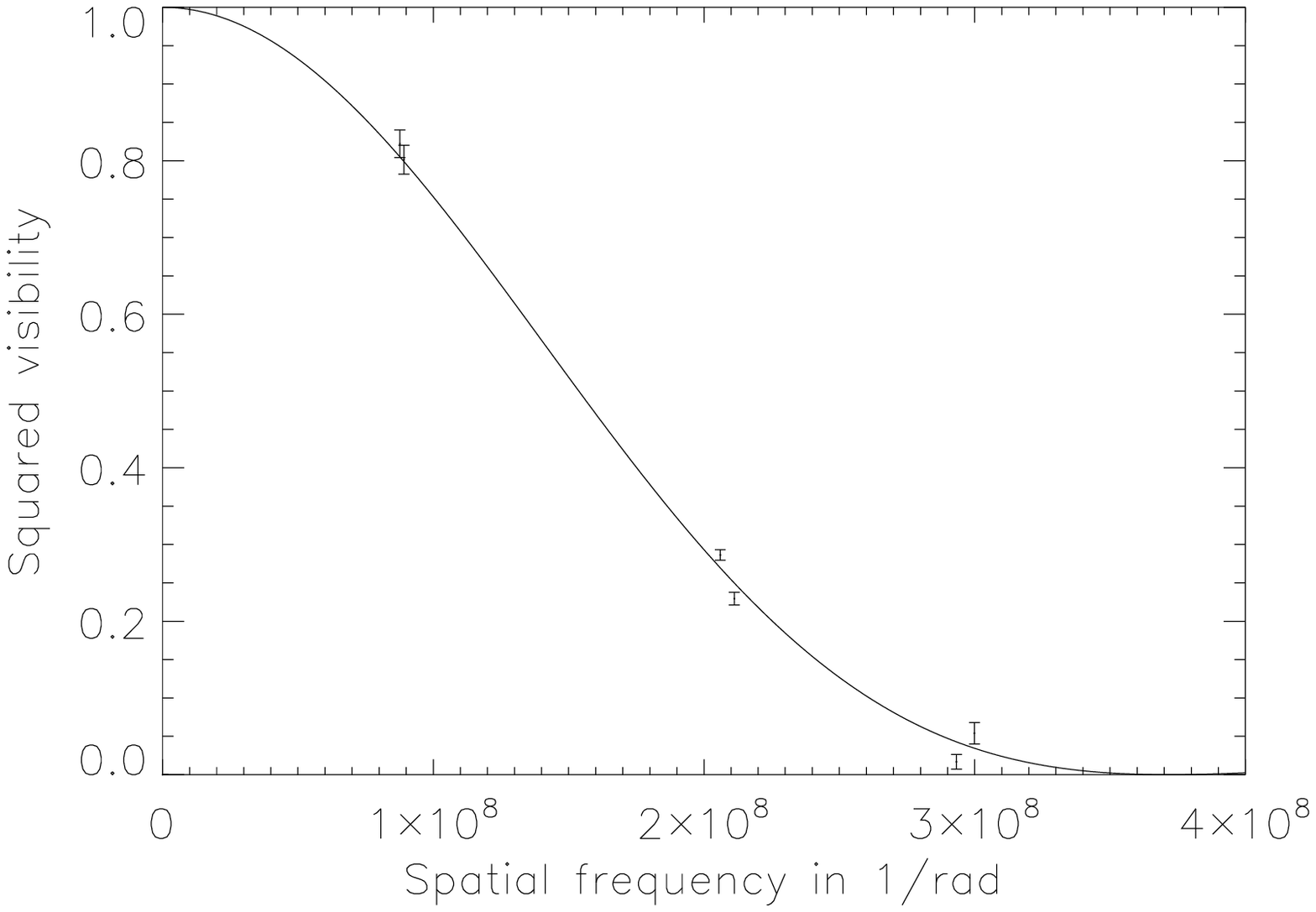}}
    \subfloat{\includegraphics[scale=0.5]{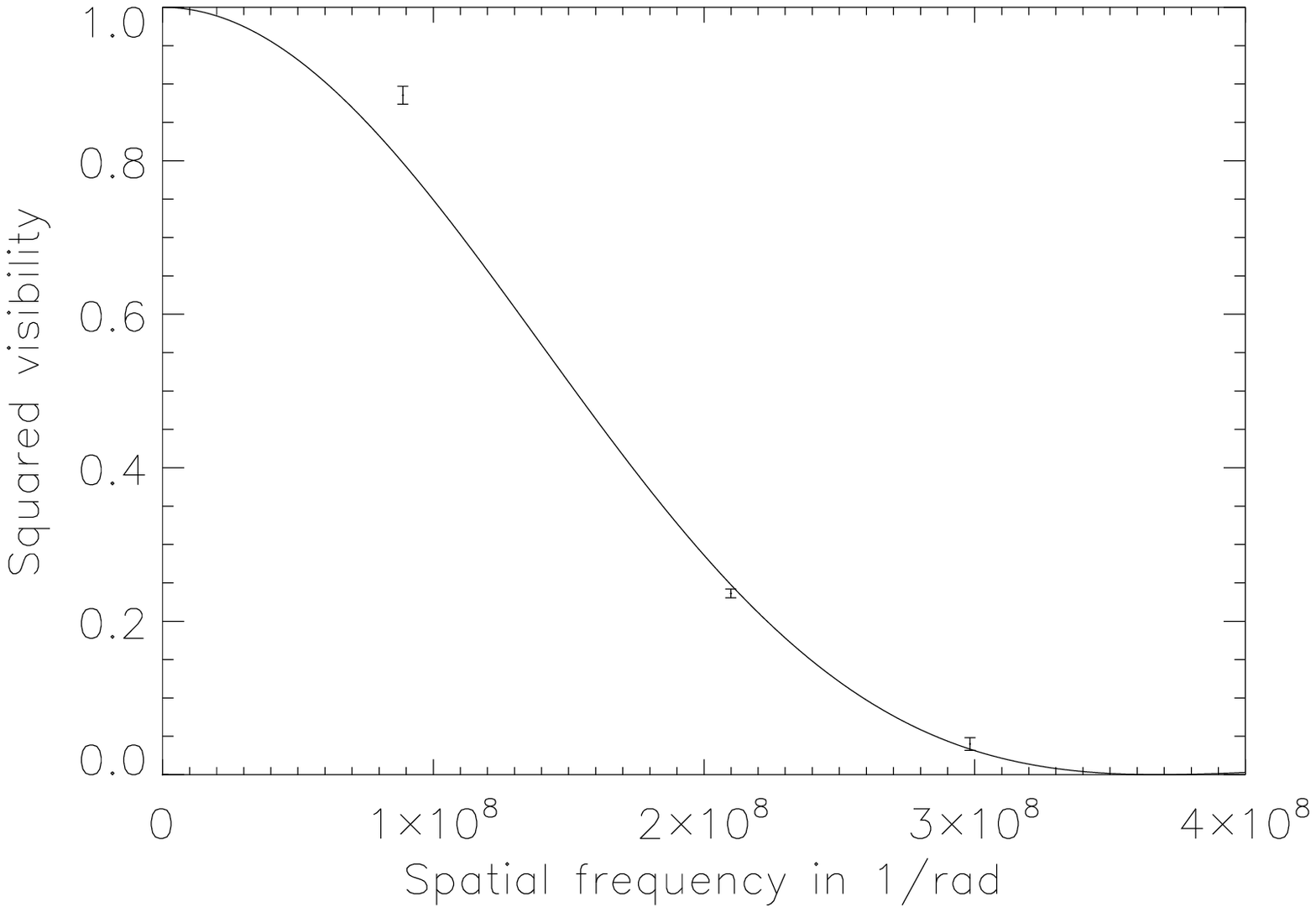}}
    \caption{Model of squared visibility for the RJD 55826.67 (left, $\theta_{\rm LD}=0.721\pm0.009$) and 55805.75 (right, $\theta_{\rm LD}=0.727\pm0.010$) obtained by LITpro.}
\end{figure}

\begin{figure*}[h]
    \subfloat{\includegraphics[scale=0.5]{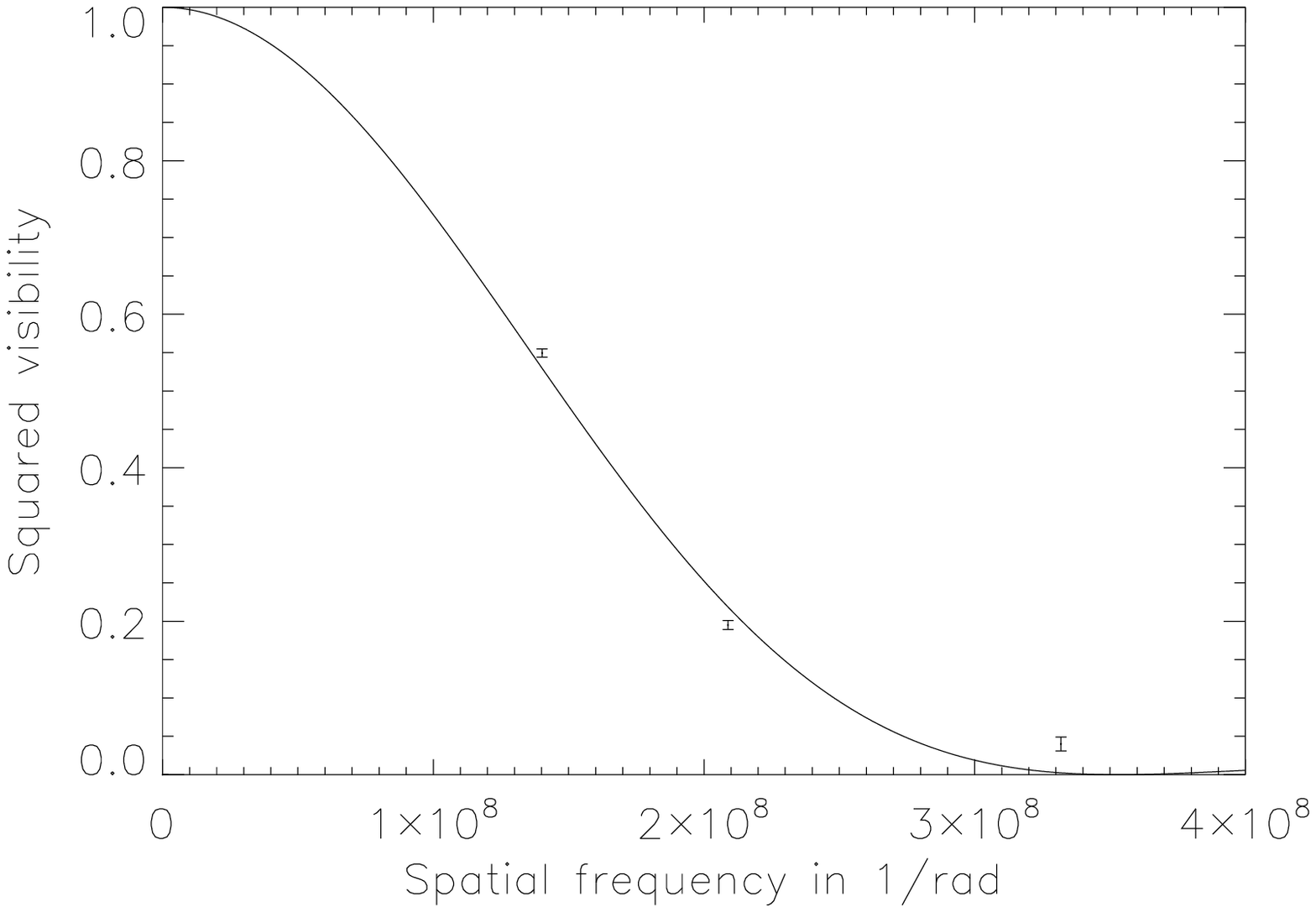}}
    \subfloat{\includegraphics[scale=0.5]{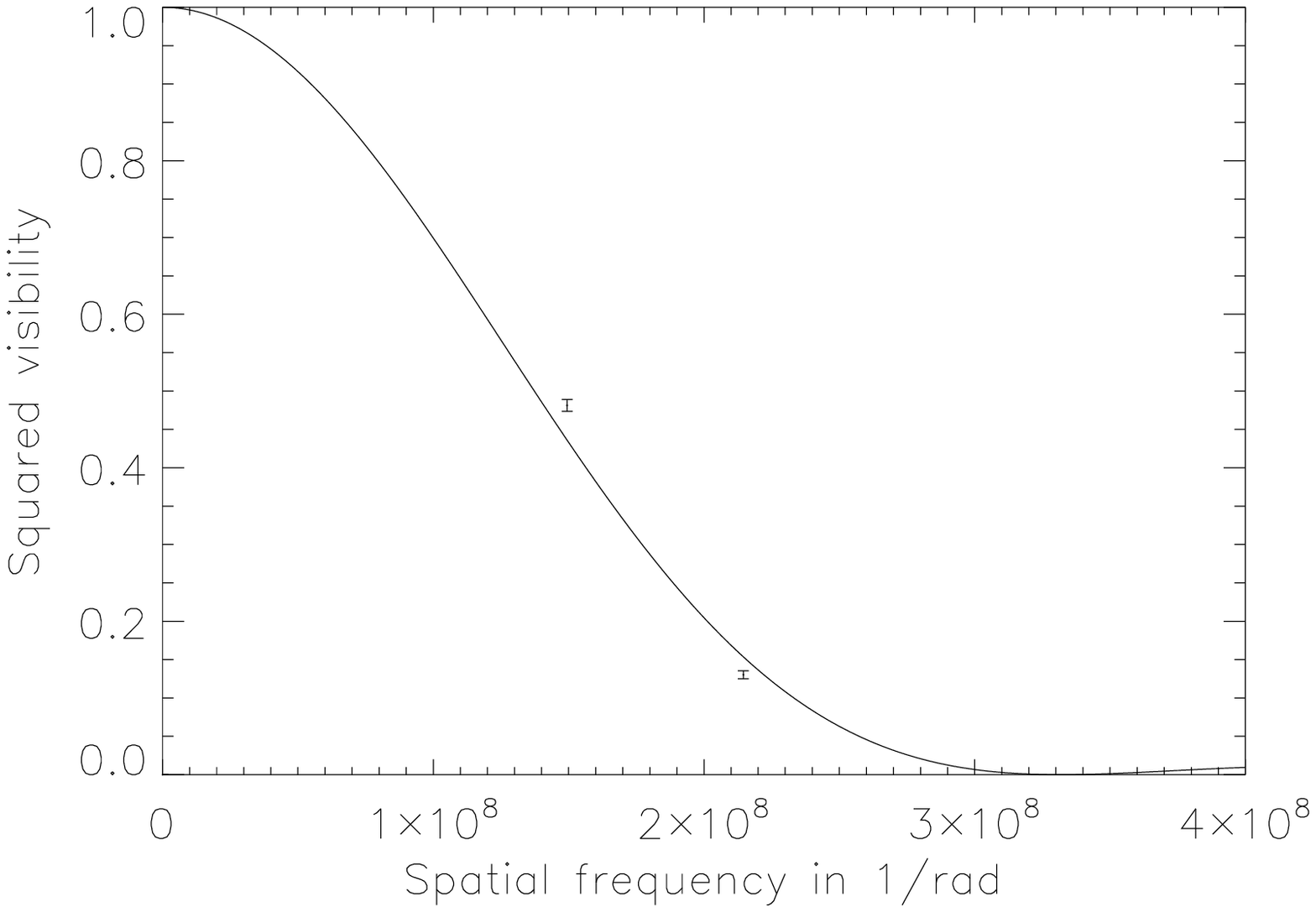}}
    \caption{Model of squared visibility for the RJD 55803.77 (left, $\theta_{\rm LD}=0.759\pm0.008$) and 55774.73 (right, $\theta_{\rm LD}=0.807\pm0.010$) obtained by LITpro.}
\end{figure*}

\begin{figure*}[h]
    \subfloat{\includegraphics[scale=0.5]{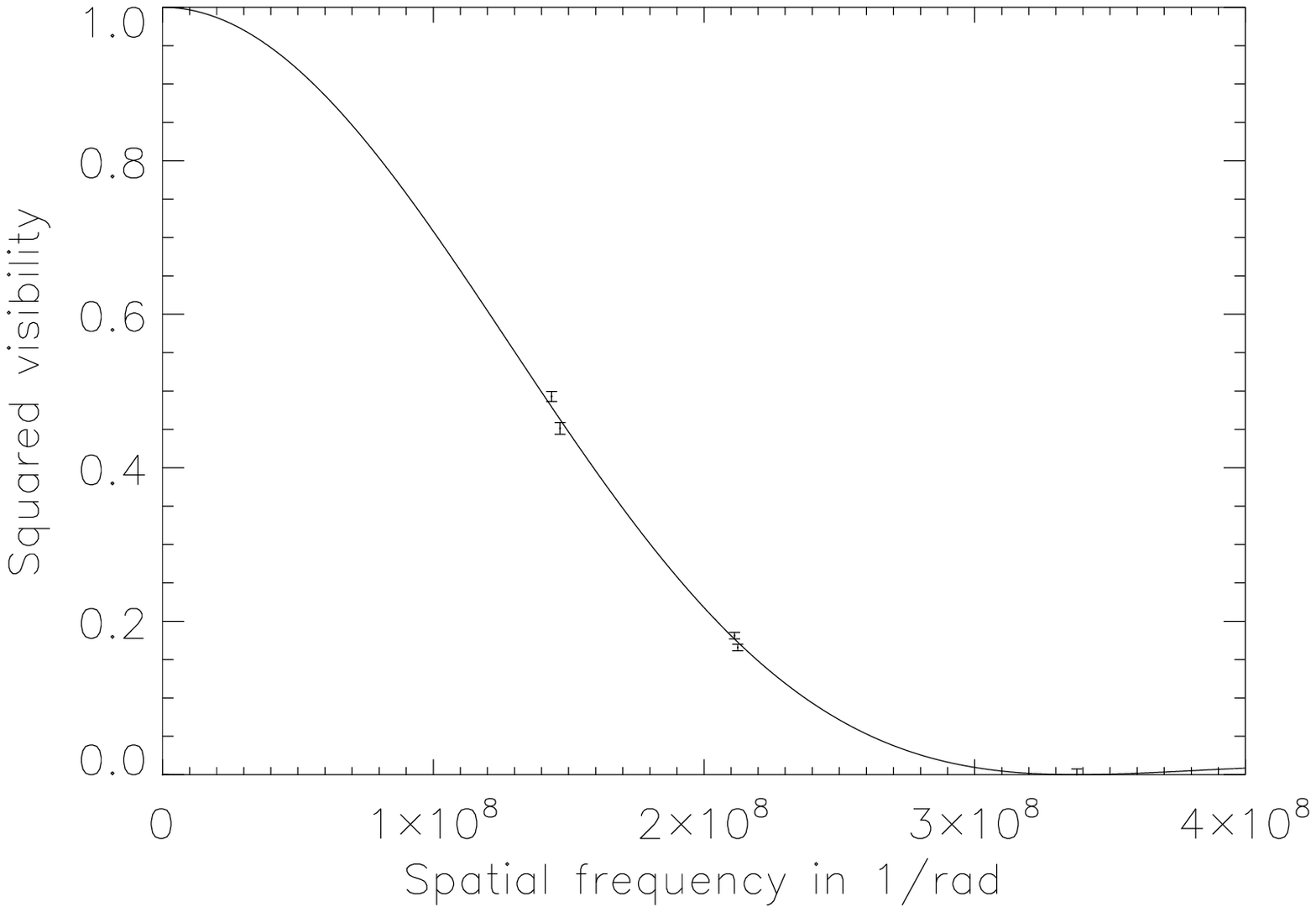}}
    \subfloat{\includegraphics[scale=0.5]{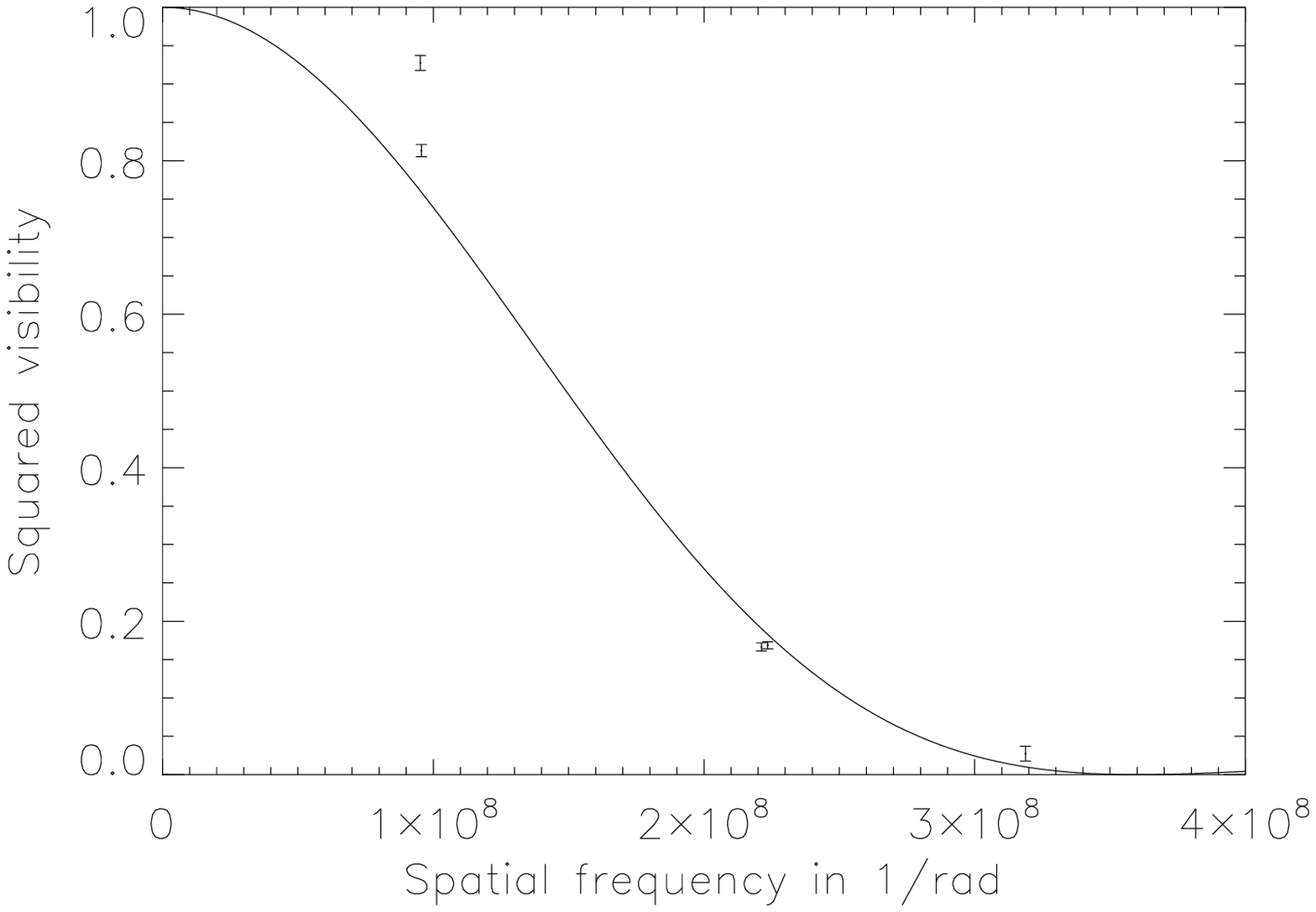}}
    \caption{Model of squared visibility for the RJD 55722.93 (left, $\theta_{\rm LD}=0.793\pm0.006$) and 55486.71 (right, $\theta_{\rm LD}=0.744\pm0.007$) obtained by LITpro.}
\end{figure*}

\begin{figure*}[h]
    \subfloat{\includegraphics[scale=0.5]{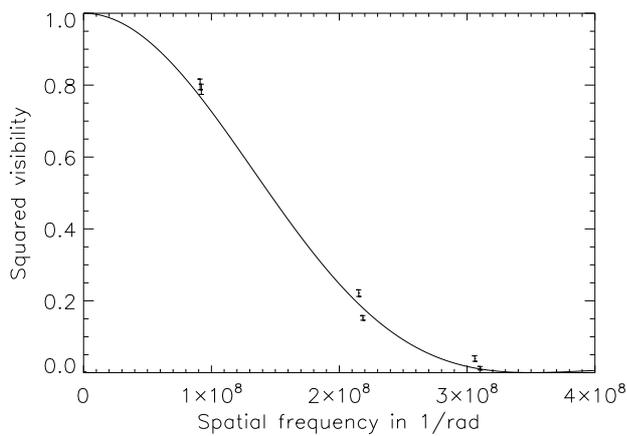}}
    \caption{Model of squared visibility for the RJD 55370.92 ($\theta_{\rm LD}=0.764\pm0.010$ obtained by LITpro.}
\end{figure*}

\end{appendix}

\end{document}